\documentclass[twocolumn, showpacs, amsmath, amssymb, aps, pra]{revtex4-1}
\usepackage{graphicx}
\usepackage{dcolumn}
\usepackage{bm}

\begin{document}

\title{Casimir expulsion of shifted configurations}

\author{Evgeny\,G.\,Fateev}
 \email{e.g.fateev@gmail.com}
\affiliation{%
Institute of mechanics, Ural Branch of the RAS, Izhevsk 426067, Russia
}%
\date{\today}

\begin{abstract}
The shift of nanosized metal configurations relative to one another can lead 
to an increase in the range of optimal lengths of wings and angles of the 
opening of cavities, at which noncompensated Casimir forces in them are 
maximal. The possibility of the existence of the effect of Casimir expulsion 
is demonstrated with a trapezoid cavity with shifted wings which are opened 
at different angles. It is also shown that shift in parallel structures 
leads to the appearance of forces tending to bring the configuration back to 
the situation before the shift. In some conditions, in systems with shifted 
structures there are moments of torsion and complex oscillation processes.
\end{abstract}

\maketitle

In Refs. \cite{Fateev:2012a, Fateev:2012b} the possibility in principle is shown that 
noncompensated Casimir force can exist in open nanosized metal cavities. The 
effect is theoretically demonstrated with a single trapezoid configuration. 
The force manifests itself as time-constant expulsion of open cavities in 
the direction of their least opening. Optimal parameters have been found for 
the angles of opening (broadening) of the cavities' generetrices and their 
lengths, at which the expulsion force is maximal. Note that this force 
significantly differs from repulsion forces capable of creating only the 
levitation-type effects over bodies-partners \cite{Jaffe:2005, 
Leonhardt:2007, Levin:2010, Rahi:2010, Rahi:2011}. 
A particular case of such configurations is classical 
Casimir parallel mirrors \cite{Casimir:1948, Casimir:1949, Milton:2001, Bordag:2009,
Bordag:2009a}. It has been found that the ends of metal mirrors act as Casimir expulsion 
forces parallel to surfaces, which make 1/5 of the specific Casimir pressure 
\cite{Leonhardt:2007}. However, these forces are completely compensated, 
i.e. in the entire configuration of parallel mirrors the effective force of 
expulsion does not exist in either direction. An interesting question arises 
whether it is possible that the balance of the forces of expulsion will 
change in trapezoid and parallel configurations when the planes shift 
relative to one another. The possibility of the evaluation of the influence 
of the planes' shift \cite{Gies:2006} upon the transverse Casimir force 
is also important. 

\begin{center}
\textbf{Theory}
\end{center}

Let us consider a configuration with trapezoid cavities with shifted wings 
using it for investigating the possibility of the existence of Casimir 
expulsion forces. Note that each single cavity should be understood as an 
open thin-walled metal shell composed of metal flat plates (wings) with one 
or several outlets. Inner and outer surfaces of the cavity should have the 
properties of perfect mirrors. The cavity can completely be immersed into a 
material medium or be its part with the dielectric permeability properties 
differing from those of physical vacuum.

Let us present a single trapezoid configuration in Cartesian coordinates as 
two thin metal plates with the surface width $L$ (oriented along the $z$ 
axis) and length $R$ (a wing) situated at the distance $a$ from one another; 
the angle $\varphi $ of the cavities' opening between the plates can be changed 
(simultaneously for both wings of the trapezoid cavity by the same value) as 
it is shown in Fig.\hyperlink{fig1} 1. Let us also take into account the possibility of the 
shift of one of the cavity wings in the direction opposite to the $x$ axis 
by step $\Delta x$. For the given problem geometry, the angle $\varphi $ should 
not exceed the value $\varphi \leqslant \mbox{arccot}\left( {{\Delta x} 
\mathord{\left/ {\vphantom {{\Delta x} a}} \right. 
\kern-\nulldelimiterspace} a} \right)$, otherwise the situation can occur 
when the configuration of cavities with two neighboring cavities is formed. 
This situation requires somewhat different problem statement.
\begin{figure*}[htbp]
\hypertarget{fig1}
\centerline{
\includegraphics[width=2.6in,height=2.2in]{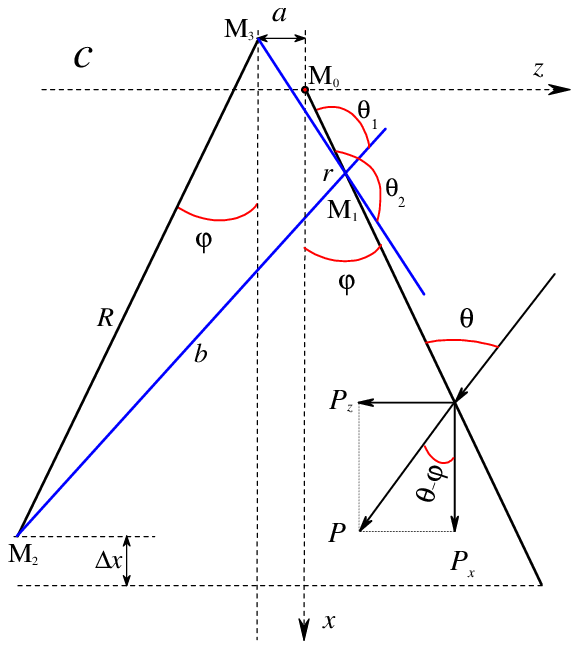}
\includegraphics[width=2.4in,height=2.1in]{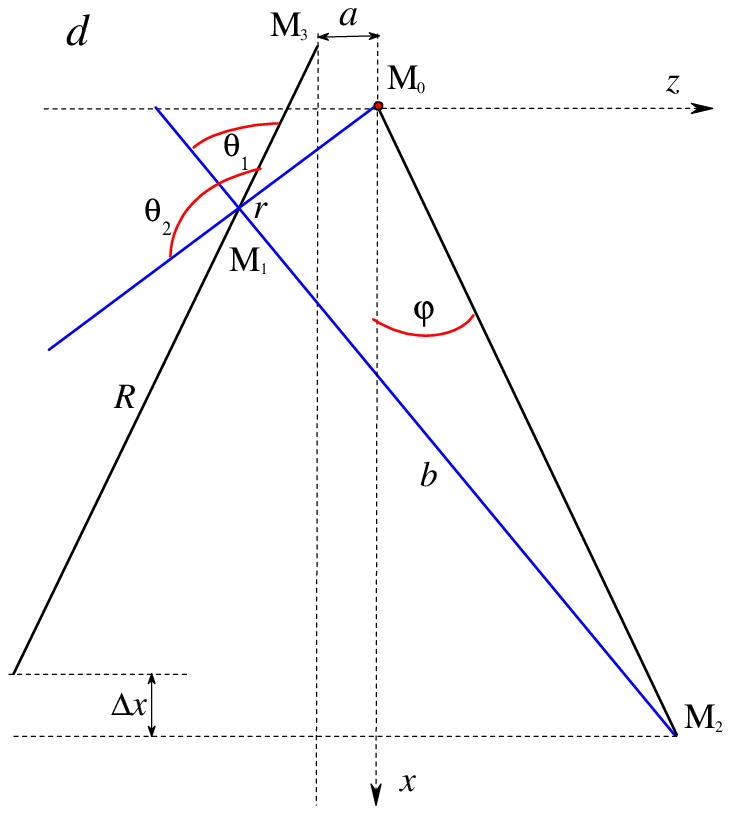}}
\label{fig1}
\caption{$c$ - schematic view of the configuration of the symmetric trapezoid 
cavity with the left wing (with the length $R$) shifted by step $\Delta x$ 
against the $x$ axis, a particular case of which are parallel wings (for 
$\varphi =0$) and a triangle (at $a=0$ and $\Delta x=0$). The cavity with the 
width $L$ in the $y$ direction is shown in the Cartesian coordinates ($x,z$). 
Blue straight lines designate virtual rays with length $b$ going from a 
point $\mbox{M}_1 $ under the limit angle $\Theta _1 $ and $\Theta _2 $ to 
the right surface of the cavity and ending at the ends of the opposite 
cavity wing at points $\mbox{M}_2 $ and $\mbox{M}_3 $, respectively; $d$ - a 
scheme of the rays going under the limit angles $\Theta _1 $ and $\Theta _2 
$ from a random point $\mbox{M}_1 $ on the left wing to the ends of the 
right wing.}
\end{figure*}
Note that further the concept developed in Ref. \cite{Fateev:2012a} is 
fully applicable for the above cavities. Thus, for one cavity wing, in the 
first approximation the expulsion force in the $x$ direction can be found in 
the form \cite{Fateev:2012a}
\begin{equation}
\label{eq1}
F_x =\int\limits_0^L {dy} \int\limits_0^R {P_x (\varphi ,\Theta ,r)} \,dr.
\end{equation}
Here the local specific force of expulsion at every point$r$on the cavity 
wing with the length $R$ and width $L$ is
\begin{equation}
\label{eq2}
\begin{array}{c}
 P_x =-\frac{\hbar c\pi ^2}{240 s^4}\int\limits_{\Theta _1 }^{\Theta _2 } 
\sin (\Theta -2\varphi )^4\cos (\Theta -\varphi )d\Theta \\ 
 =-\frac{\hbar c\pi ^2}{240 s^4}A_2 (\varphi ,\Theta _1 ,\Theta _2 ), \\ 
 \end{array}
\end{equation}
where 
\begin{equation}
\label{eq3}
\begin{array}{l}
A_2 (\varphi ,\Theta _1 ,\Theta _2 ) \\ 
 =\frac{1}{240}\Bigl[ 90\sin (\varphi -\Theta _1 )-90\sin (\varphi -\Theta _2 ) \\ 
 +60\sin (3\varphi -\Theta _2 )-60\sin (3\varphi -\Theta _1 ) \\ 
 +20\sin (5\varphi -3\Theta _2 )-20\sin (5\varphi -3\Theta _1 ) \\ 
 +5\sin (7\varphi -3\Theta _1 )-5\sin (7\varphi -3\Theta _2 ) \\ 
 +3\sin (9\varphi -5\Theta _1 )-3\sin (9\varphi -5\Theta _2 )\Bigr].
 \end{array}
\end{equation}
In formula (\ref{eq2}) $\hbar =h/2\pi $ is reduced Planck constant, $c$ is light 
velocity; functional expressions for the limit angles $\Theta _1 
,\;\Theta _2 $ in the trapezoid cavity and the parameter $s$ are as follows 
\begin{equation}
\label{eq4}
\Theta _1 =\mbox{arccos}\left(  {-\frac{r+a\sin \varphi -R\cos 2\varphi}{\sqrt 
{\begin{array}{l}
 \left( {a+R\sin \varphi + r\sin \varphi} \right)^2 \\ 
 +\left( {r\cos \varphi - R\cos \varphi} \right)^2 \\ 
 \end{array}} }} \right),
 \end{equation}
\begin{equation}
\label{eq5}
\Theta _2 =\mbox{arccos}\left[ {-\frac{r+a\sin \varphi}{\sqrt 
{a^2+r^2+2r(a\sin \varphi} )}} \right],
\end{equation}
\begin{equation}
\label{eq6}
s=\frac{\sin (2\varphi -\Theta _2 )(a+r\sin \varphi)}{\sin (\varphi -\Theta _2)}.
\end{equation}
For the trapezoid figure the Casimir force in the $z$ direction can be found 
in the form \cite{Fateev:2012a} 
\begin{equation}
\label{eq7}
F_z =\int\limits_0^L {dy} \int\limits_0^R {P_z (\varphi ,\Theta ,r)} \,dr.
\end{equation}
Here, at every point $r$ on the cavity wing the specific Casimir force in 
the $z$ direction is
\begin{equation}
\label{eq8}
\begin{array}{c}
 P_z (r)=-\frac{\hbar c\pi ^2}{240 s^4}\int\limits_{\Theta _1 }^{\Theta _2 } 
\sin (\Theta -2\varphi )^4\sin (\Theta -\varphi )d\Theta \\ 
 =-\frac{\hbar c\pi ^2}{240 s^4}A_1 (\varphi ,\Theta _1 ,\Theta _2 ), \\ 
 \end{array}
\end{equation}
where 
\begin{equation}
\label{eq9}
\begin{array}{l}
 A_1 (\varphi ,\Theta _1 ,\Theta _2 ) \\ 
 =\frac{1}{240}\Bigl[ 90\cos (\varphi -\Theta _1 )-90\cos (\varphi -\Theta _2 ) \\ 
 +60\cos (3\varphi -\Theta _2 )-60\cos (3\varphi -\Theta _1 ) \\ 
 +20\cos (5\varphi -3\Theta _2 )-20\cos (5\varphi -3\Theta _1 ) \\ 
 +5\cos (7\varphi -3\Theta _1 )-5\cos (7\varphi -3\Theta _2 ) \\ 
 +3\cos (9\varphi -5\Theta _1 )-3\cos (9\varphi -5\Theta _2 ) \Bigr]. 
\end{array}
\end{equation}
To take into account the possibility of the influence of the left cavity 
wing shifts $\Delta x$ relative to the right wing on the Casimir forces, it 
is necessary to find new functional expressions for the limit angles $\Theta 
_1 ,\;\Theta _2 $. Let us use the presentation of the directing vectors 
corresponding to the rays for the limit angles $\Theta _1 ,\;\Theta _2 $ and 
the generatrix of the right cavity wing on the scheme of the trapezoid 
configuration (Fig.\hyperlink{fig1} 1$c)$. Let us designate the point data on the plan of the 
trapezoid cavity: $M_0 (x_0 ,z_0 )$, $M_1 (x_1 ,z_1 )$, $M_2 (x_2 ,z_2 )$ and 
$M_3 (x_3 ,z_3 )$. Then the vector $\overrightarrow {M_0 M} _1 =(x_1 -x_0 
;z_1 -z_0 )$ can be chosen as a directing vector of the generatrix of the 
right cavity wing. $\overrightarrow {M_1 M} _2 =(x_2 -x_1 ;z_2 -z_1 )$ is 
chosen as a directing vector of the straight line $b$. $\overrightarrow {M_1 
M} _3 =(x_3 -x_1 ;z_3 -z_1 )$ is chosen as a vector of the straight line 
between the extreme upper point of the left wing $M_3 (x_3 ,z_3 )$ and the 
point $M_1 (x_1 ,z_1 )$. The corresponding coordinates can be written in the 
form: $x_0 =0$; $z_0 =0$; $x_1 =r\cos \varphi $; $z_1 =r\sin \varphi $; $x_2 
=R\cos \varphi -\Delta x$; $z_2 =-R\sin \varphi -a$; $x_3 =-\Delta x$; $z_3 =-a$. 
The cosines of the angles between the directrices are determined as follows
\[
\begin{array}{l}
 \cos \Theta _{_1 }^{right} =\mp \frac{\overrightarrow {M_0 M} _1 
\cdot \overrightarrow {M_1 M} _2 }{\left\| {\overrightarrow {M_0 M} _1 } 
\right\|\cdot \left\| {\overrightarrow {M_1 M} _2 } \right\|} \\ 
 =\pm \frac{(x_1 -x_0 )(x_2 -x_1 )+(z_1 -z_0 )(z_2 -z_1 )}{\sqrt {(x_1 -x_0 
)^2+(z_1 -z_0 )^2} \sqrt {(x_2 -x_1 )^2+(z_2 -z_1 )^2} }, \\ 
 \\ 
 \end{array}
\]
and
\[
\begin{array}{l}
 \cos \Theta _{_2 }^{right} =\mp \frac{\overrightarrow {M_0 M} _1 
\cdot \overrightarrow {M_1 M} _3 }{\left\| {\overrightarrow {M_0 M} _1 } 
\right\|\cdot \left\| {\overrightarrow {M_1 M} _3 } \right\|} \\ 
 =\pm \frac{(x_1 -x_0 )(x_3 -x_1 )+(z_1 -z_0 )(z_3 -z_1 )}{\sqrt {(x_1 -x_0 
)^2+(z_1 -z_0 )^2} \sqrt {(x_3 -x_1 )^2+(z_3 -z_1 )^2} }. \\ 
 \end{array}
\]
Now let us find the angles $\Theta_{1}^{right}$ and 
$\Theta_{2}^{right}$ in the form
\[
\begin{array}{lcl}
\Theta_{1}^{right}\\
=\arccos \left[ {\frac{r+a\sin \varphi 
+\Delta x \cos \varphi - R\cos 2\varphi }
{-\sqrt {\left[ {a+(R+r)\sin \varphi } \right]^2 
 + \left[ {\Delta x+(r-R)\cos \varphi} \right] ^{2}}}}\right] 
\end{array},\eqno(10)
\]
$$
\begin{array}{lcl}
\Theta_{2}^{right}\\
=\arccos \left[ {\frac{r+a\sin \varphi 
+\Delta x\cos \varphi }
{-\sqrt {a^2+\Delta x^2+r^2+2a\,r\sin \varphi
+2\Delta x{\kern 1pt}r\cos \varphi } }} \right]\end{array}.\eqno(11) 
$$
\begin{figure}[htbp]
\hypertarget{fig2}
\centerline{
\includegraphics[width=1.5in,height=1.5in]{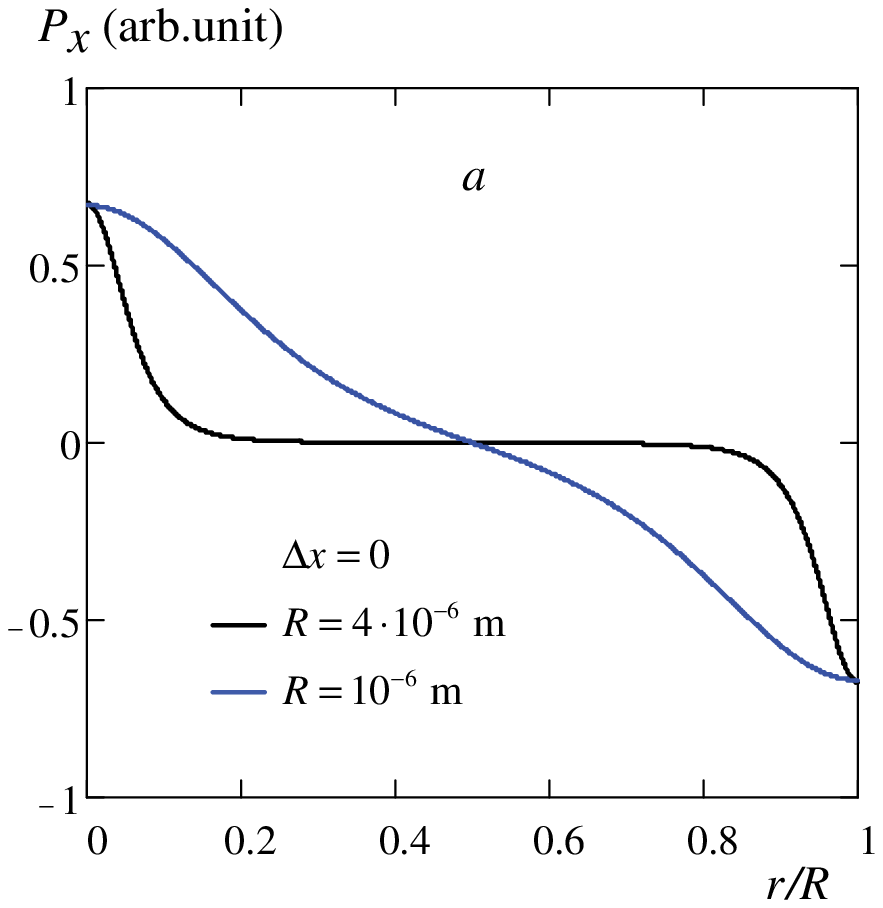}
\includegraphics[width=1.5in,height=1.5in]{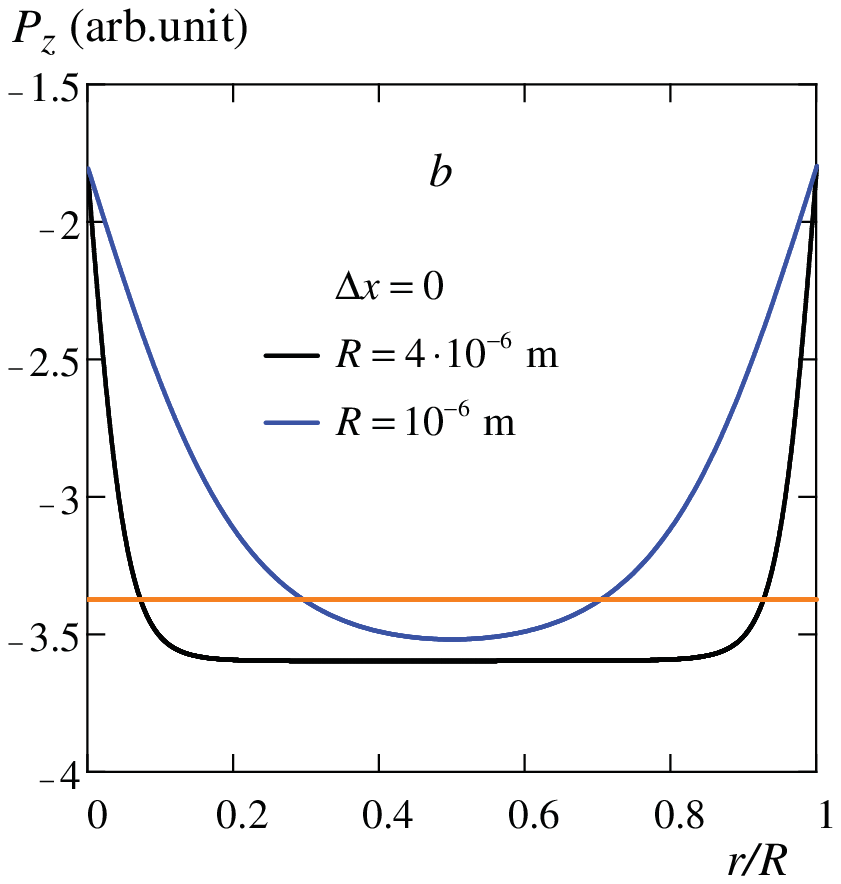}\linebreak }
\centerline{
\includegraphics[width=1.5in,height=1.5in]{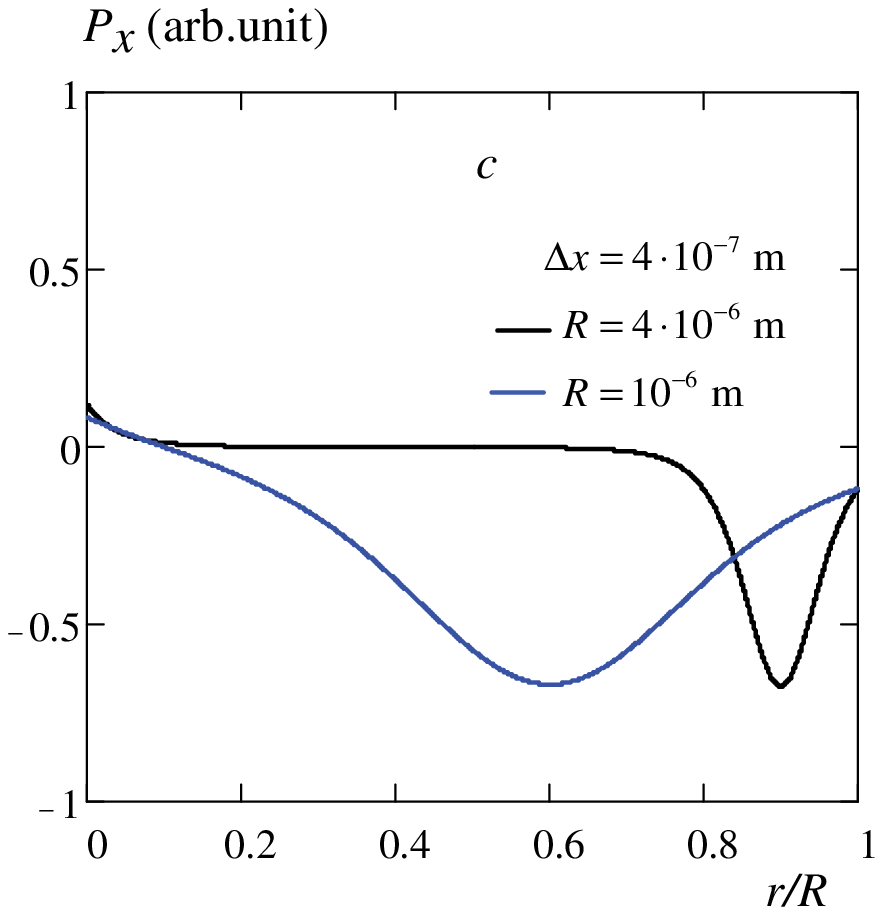}
\includegraphics[width=1.5in,height=1.5in]{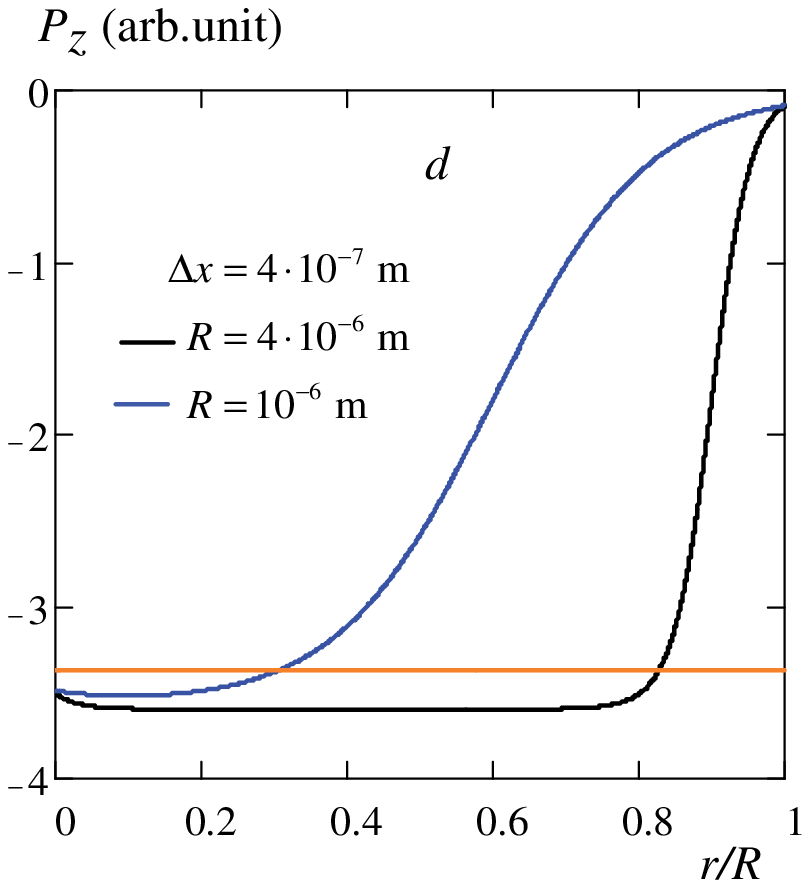}\linebreak }
\centerline{
\includegraphics[width=1.5in,height=1.5in]{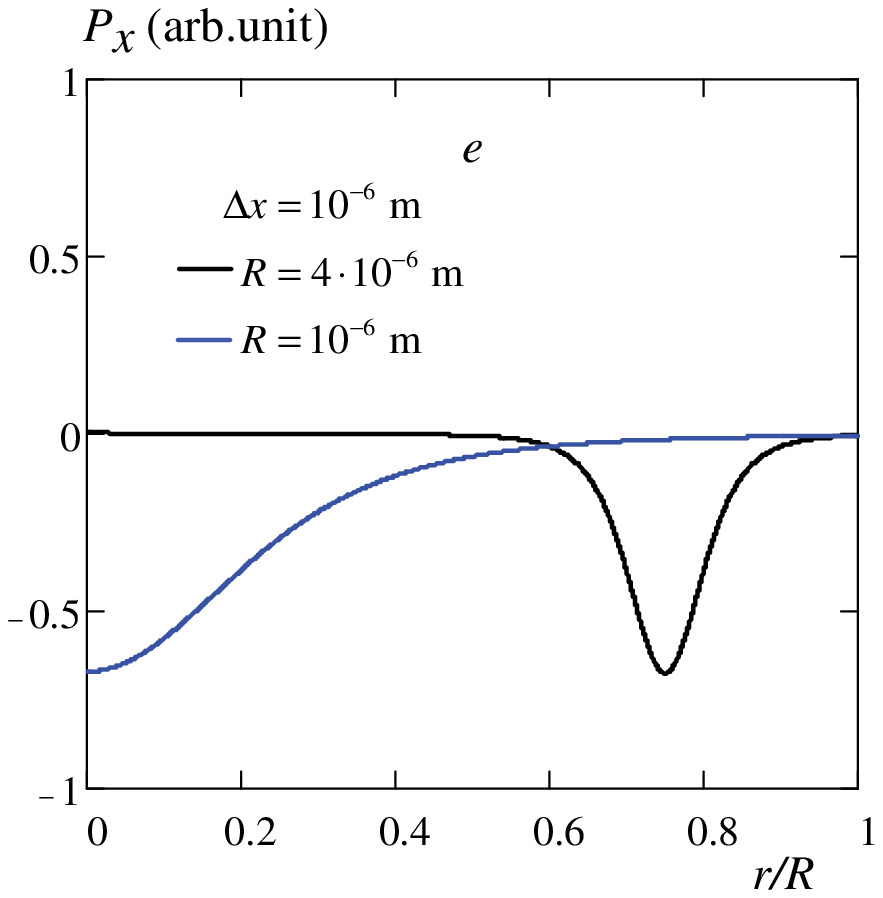}
\includegraphics[width=1.5in,height=1.5in]{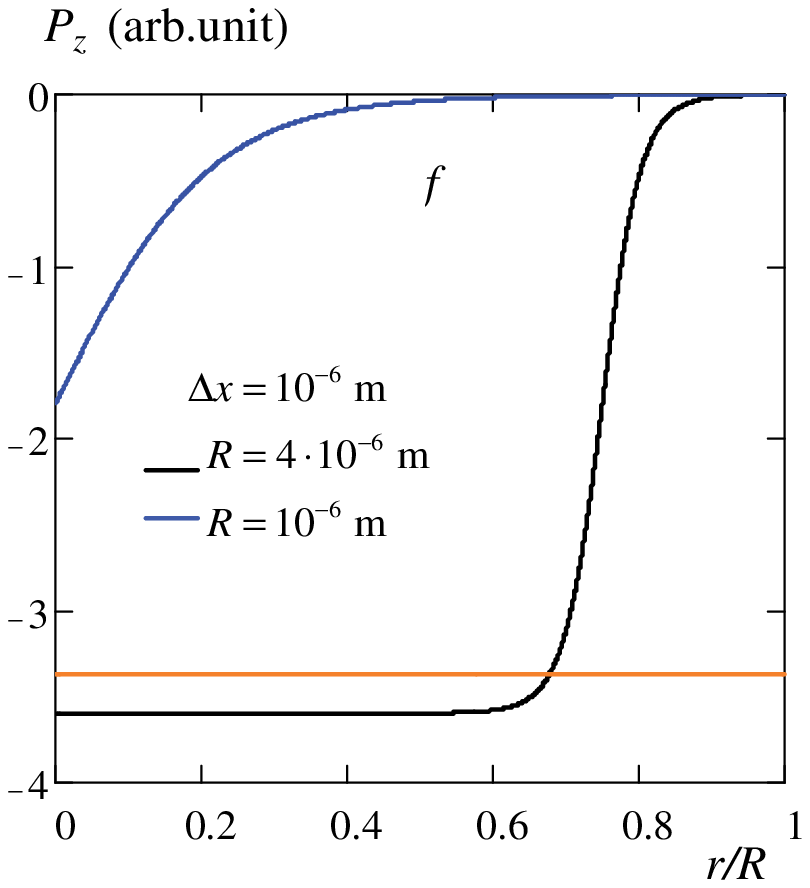}\linebreak }
\label{fig2}
\caption{Specific local forces of expulsion ($a)$ and pressure ($b)$ along the $x$ 
axis for strictly parallel wings ($\varphi =0)$ in the absence of shift. 
Respectively, local forces ($c)$ and ($d)$ at the shift of the cavity left wing 
for $\varphi =0$ by $\Delta x=4\times 10^{-7}\;\mbox{m}$. Local specific forces 
($e)$ and ($f)$ at the shift of the cavity left wing by $\Delta 
x=10^{-6}\;\mbox{m}$. The red line ($b, d, f)$ shows the classical level of the 
Casimir pressure for $a=4\times 10^{-7}$\, m. }
\end{figure}
To obtain the integral dependencies of expulsion and pressure for both wings 
of the cavity it is also necessary to find the corresponding limit angles 
for the rays going from the arbitrary points of the left wing to the ends of 
the right wing in the process of shift. It is clear that at the shift 
relative to the right wing, the point data $M_0 (x_0 ,z_0 )$, $M_1 (x_1 ,z_1 
)$, $M_2 (x_2 ,z_2 )$ and $M_3 (x_3 ,z_3 )$ corresponding to the vectors 
$\overrightarrow {M_0 M} _1 =(x_1 -x_0 ;z_1 -z_0 )$, $\overrightarrow {M_1 
M} _2 =(x_2 -x_1 ;z_2 -z_1 )$, and $\overrightarrow {M_1 M} _3 =(x_3 -x_1 
;z_3 -z_1 )$ going from the left wing of the cavity (see Fig.\hyperlink{fig1} 1$d)$ will differ 
from those for the right wing. Let us write them down: $x_0 =0$;$z_0 =0$; 
$x_1 =r\cos \varphi -\Delta x$; $z_1 =-r\sin \varphi -a$; $x_2 =R\cos \varphi $; $z_2 
=R\sin \varphi $; $x_3 =-\Delta x$; $z_3 =-a$. The corresponding cosines of the 
angles between the directrices of these vectors have the form
$$
\begin{array}{l}
 \cos \Theta _{_1 }^{left} =\mp \frac{\overrightarrow {M_1 M} _3 
\cdot \overrightarrow {M_1 M} _2 }{\left\| {\overrightarrow {M_1 M} _3 } 
\right\|\cdot \left\| {\overrightarrow {M_1 M} _2 } \right\|} \\ 
 =\pm \frac{(x_3 -x_1 )(x_2 -x_1 )+(z_3 -z_1 )(z_2 -z_1 )}{\sqrt {(x_3 -x_1 
)^2+(z_3 -z_1 )^2} \sqrt {(x_2 -x_1 )^2+(z_2 -z_1 )^2} }, \\ 
 \end{array}$$
and
$$
\begin{array}{l}
 \cos \Theta _{_2 }^{left} =\mp \frac{\overrightarrow {M_0 M} _1 \cdot 
\overrightarrow {M_1 M} _3 }{\left\| {\overrightarrow {M_0 M} _1 } 
\right\|\cdot \left\| {\overrightarrow {M_1 M} _3 } \right\|} \\ 
 =\pm \frac{(x_1 -x_0 )(x_3 -x_1 )+(z_1 -z_0 )(z_3 -z_1 )}{\sqrt {(x_1 -x_0 
)^2+(z_1 -z_0 )^2} \sqrt {(x_3 -x_1 )^2+(z_3 -z_1 )^2} }. \\ 
 \end{array}$$
Let us find the angles $\Theta _{_1 }^{\mbox{left}} $ and $\Theta _{_2 
}^{\mbox{left}} $ in the form
$$
\begin{array}{lcl}
\Theta_{_1 }^{left}=\pi \\ 
-\arccos \left[{\frac{R+r+a\sin \varphi 
-\Delta x\cos \varphi -2R\cos ^2\varphi }{-\sqrt {\left[ {a+(R+r)\sin \varphi } 
\right]^2+\left[ {\Delta x+(R-r)\cos \varphi } \right]^2} }}\right]\end{array},\eqno (12)
$$
$$
\begin{array}{lcl}
\Theta _{_2 }^{left}\\
=\arccos \left[ {\frac{r+a\sin \varphi 
-\Delta x\cos \varphi }{-\sqrt {a^2+\Delta x^2+r^2+2a\,r\sin \varphi - 2\Delta 
xr\cos \varphi }}}\right]\end{array}.\eqno (13)
$$
In this case the change of the parameter $s$ will correspond to formula (\ref{eq5}).
Thus, the forces acting upon the entire configuration at the shift of the 
wings relative to one another (if the configuration is fixed in a new 
position after the shift) are the sum of the corresponding forces for the 
right and left wings of the cavity

$F_x =F_x^{right} +F_x^{left} $ and $F_z =F_z^{right} +F_z^{left}$.  (14)

\begin{center}
\textbf{Calculation results}
\end{center}

Using formulae (1-14) it is possible to find the character of the specific 
Casimir forces $P_z (r)$ and $P_x (r)$ along the $x$ axis for two parallel 
plates ($\varphi =0)$ having the same length $R$ and the distance $a=4\times 
10^{-7}\,\mbox{m}$ between them (see Fig.\hyperlink{fig2} 2). In Fig. \hyperlink{fig2} 2 it can be seen that 
at $R/a\geqslant 1$ on the figure boundaries the specific force $P_z (r)$ is 
always less by half than that in the centre of the configuration (Fig. \hyperlink{fig2}
2$b)$. The smaller is the wing length $R$, the more inhomogeneous are the 
forces of compression. In addition it can be seen that there are forces $P_x 
(r)$ in the configuration compressing the ends of the parallel plates toward 
their centers (Fig.\hyperlink{fig2} 2$a)$.
\begin{figure*}[htbp]
\hypertarget{fig3}
\centerline{
\includegraphics[width=1.6in,height=1.6in]{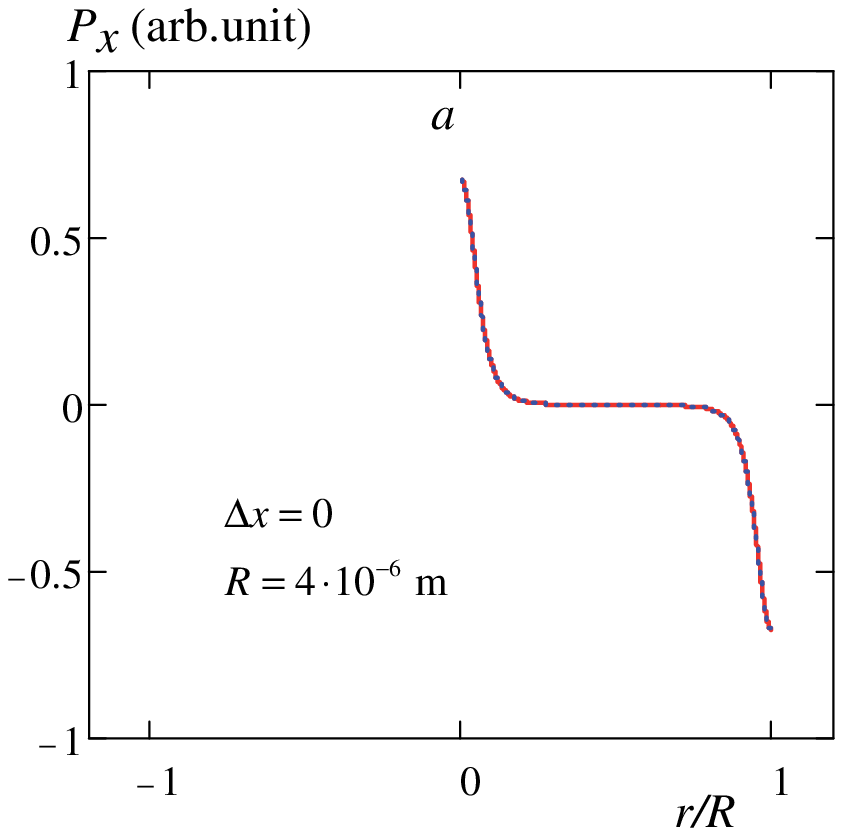}
\includegraphics[width=1.6in,height=1.6in]{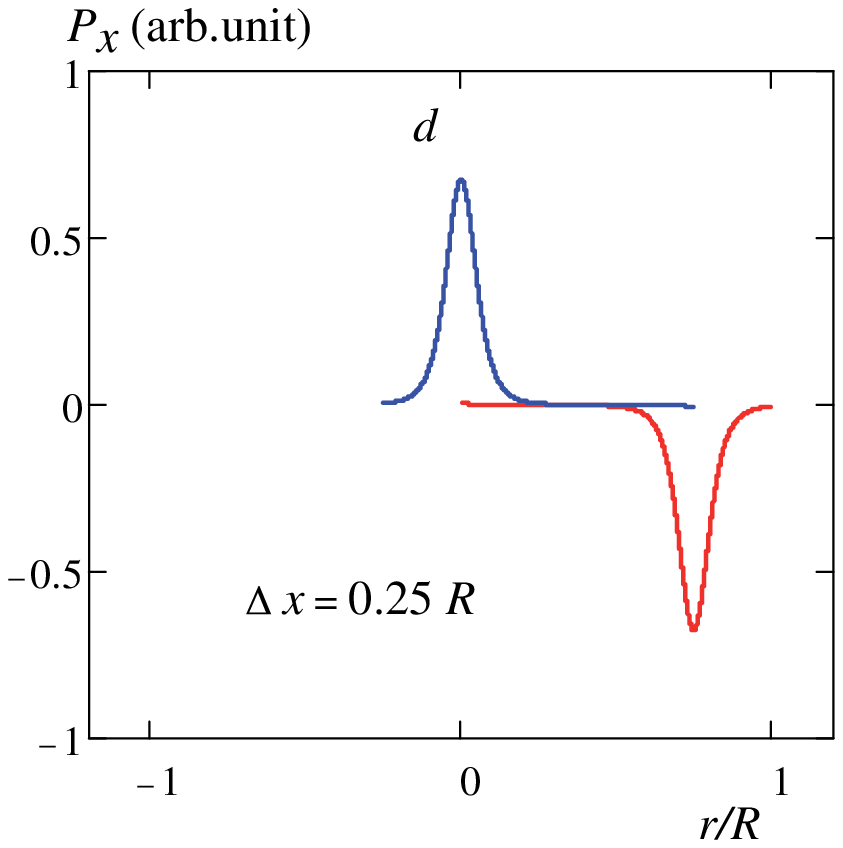}
\includegraphics[width=1.6in,height=1.6in]{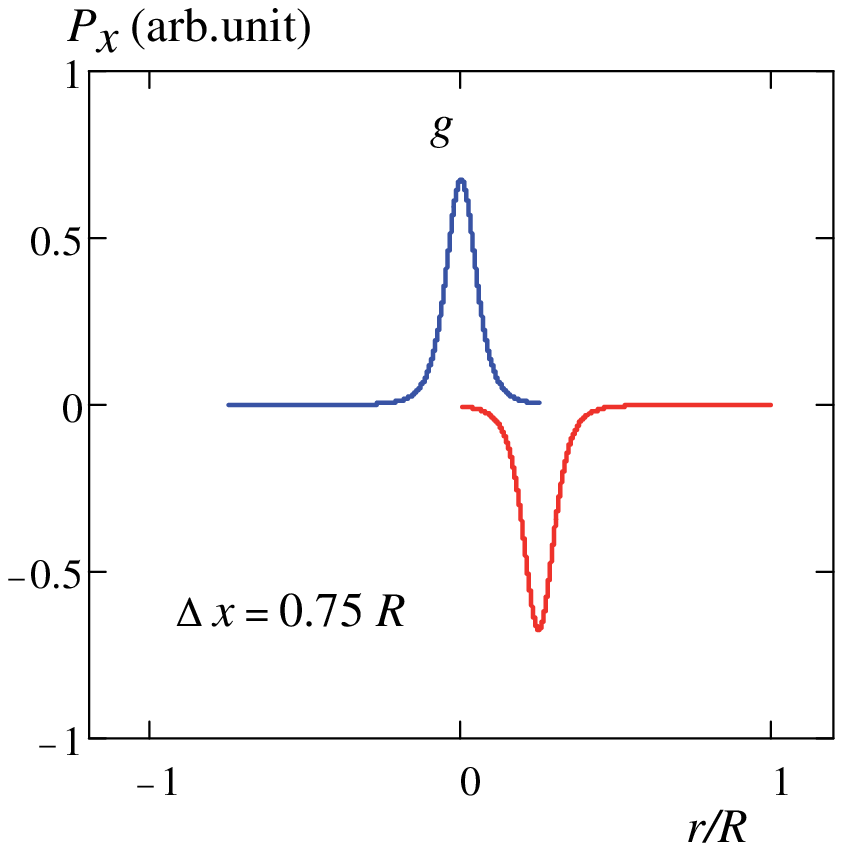}
\includegraphics[width=1.6in,height=1.6in]{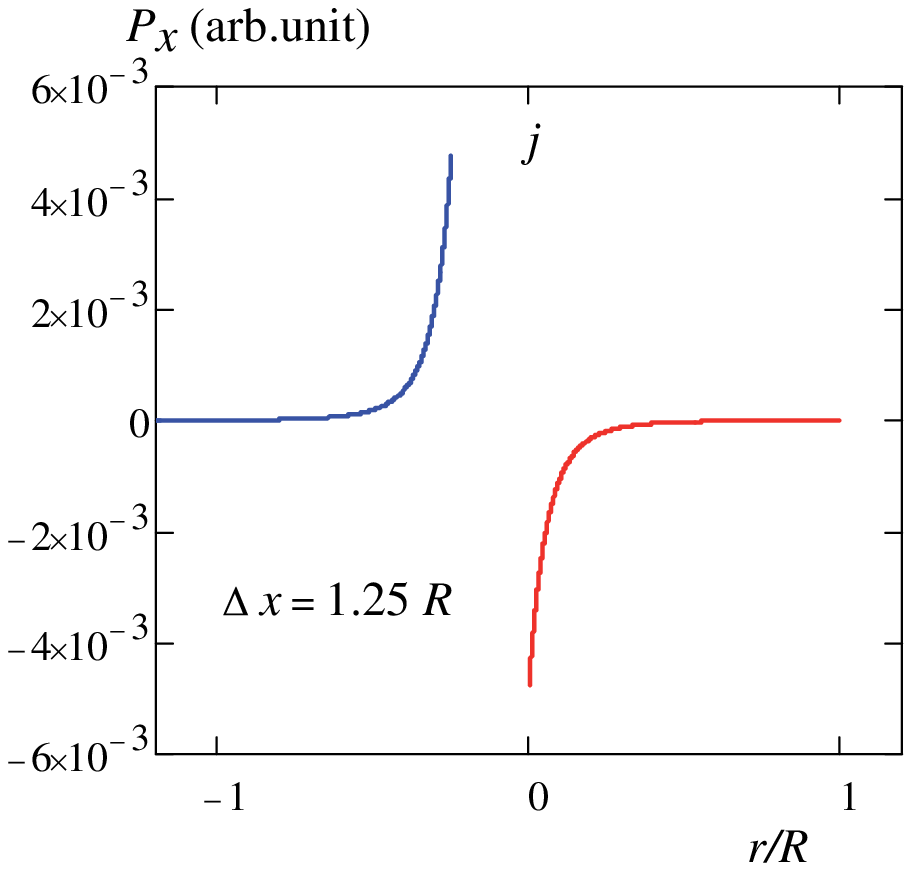}\linebreak }
\centerline{
\includegraphics[width=1.6in,height=1.6in]{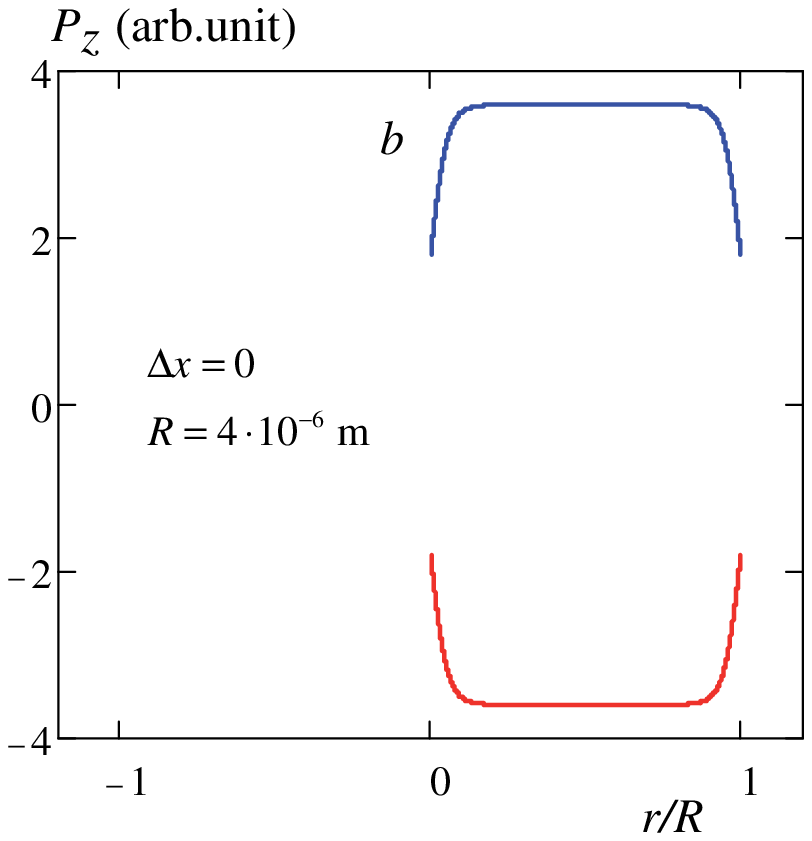}
\includegraphics[width=1.6in,height=1.6in]{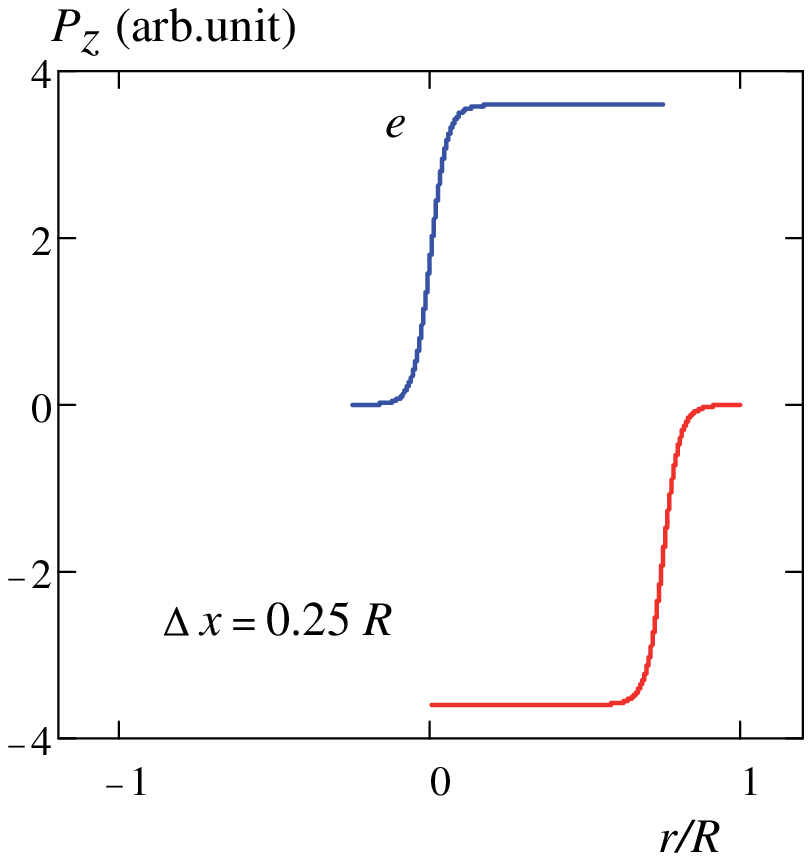}
\includegraphics[width=1.6in,height=1.6in]{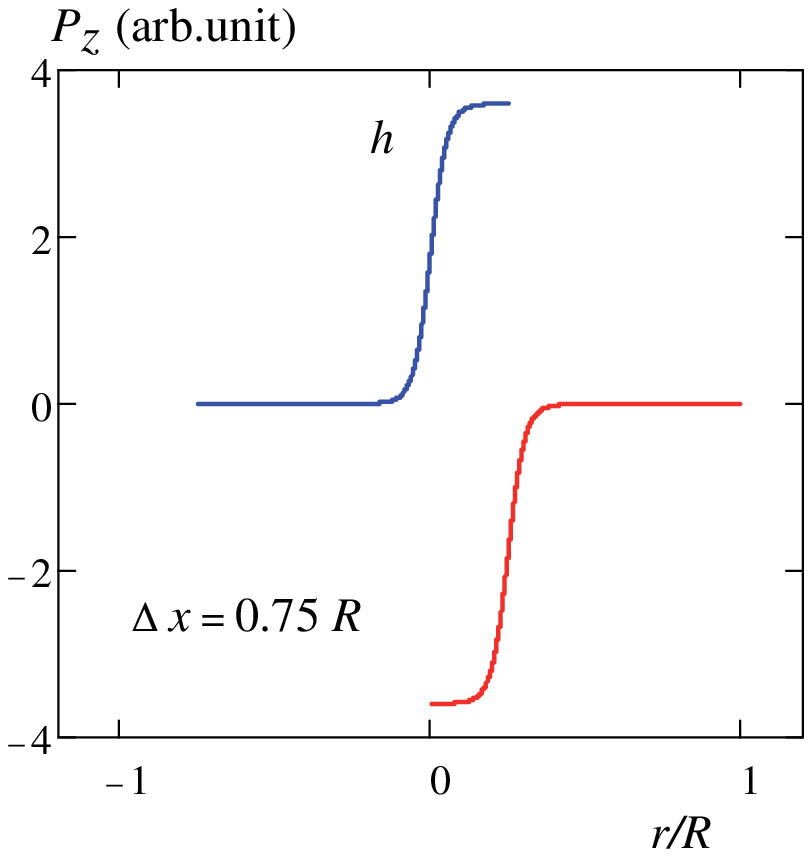}
\includegraphics[width=1.6in,height=1.6in]{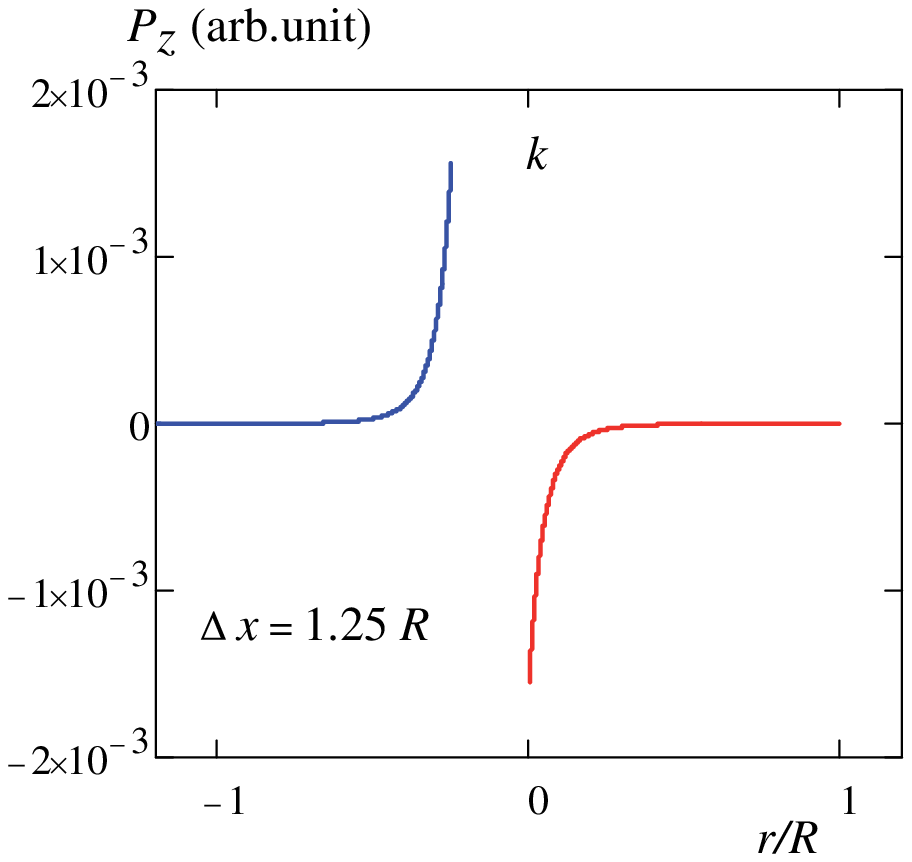}\linebreak }
\centerline{
\includegraphics[width=1.6in,height=0.8in]{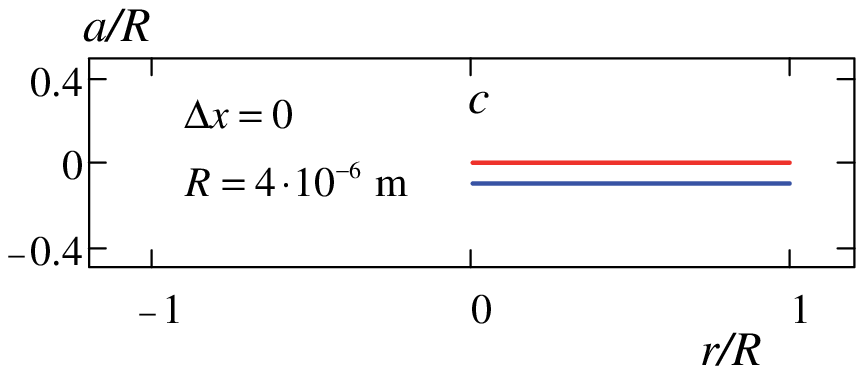}
\includegraphics[width=1.6in,height=0.8in]{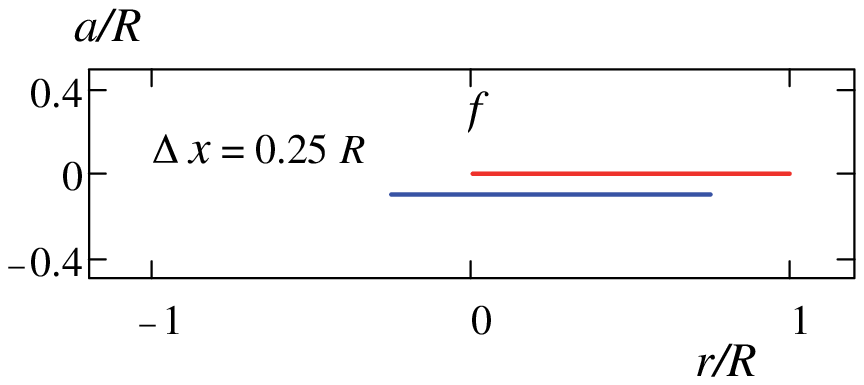}
\includegraphics[width=1.6in,height=0.8in]{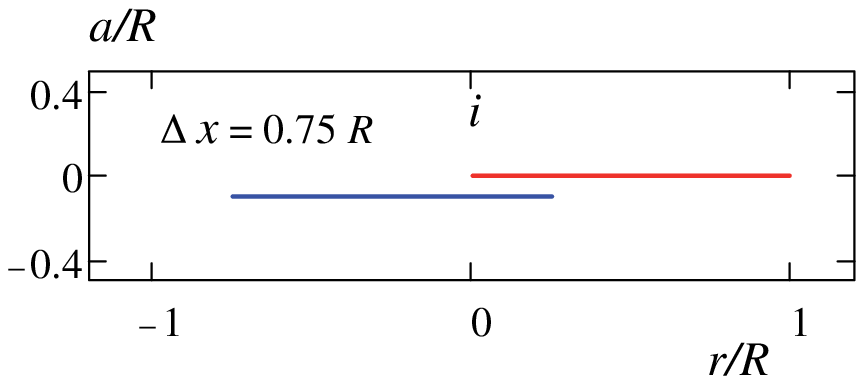}
\includegraphics[width=1.6in,height=0.8in]{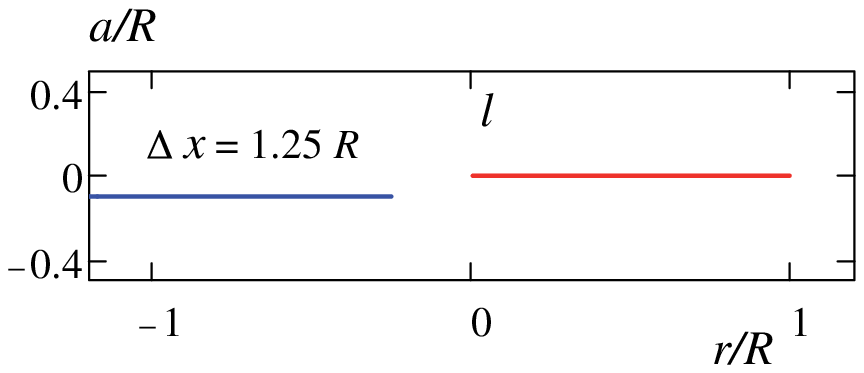}}
\label{fig3}
\caption{Specific local forces of expulsion (\textit{a, d, g, j}) and pressure
(\textit{b, e, h, k}) along the $x$ 
axis for strictly parallel plates ($\varphi =0)$ at different relative shifts 
$\Delta x/R$. In the figure, the red lines show the dependences of the local 
forces corresponding to the right plates (red) (\textit{c, f, i, l}), and the blue lines show 
the dependences corresponding to the left plates (blue).}
\end{figure*}

The specific force of compression of the two parallel plates ($\varphi =0$) 
along the $x$ axis makes up $3/8$ of the Casimir pressure on the plates 
along the $z$ axis. The smaller is the length $R$, the larger part of the 
cavity wing is subjected to the action of such forces. However, along the 
$x$ axis the integral Casimir forces $P_x (r)$ compensate one another, i.e. 
is the directed force does not act upon the entire configuration in any 
direction. There will be similar character of dependences at the rescaling 
of dimensional parameters of the configuration to any small values within 
physically reasonable limits restricted by sizes of atoms.
\begin{figure*}[htbp]
\hypertarget{fig4}
\centerline{
\includegraphics[width=1.75in,height=1.75in]{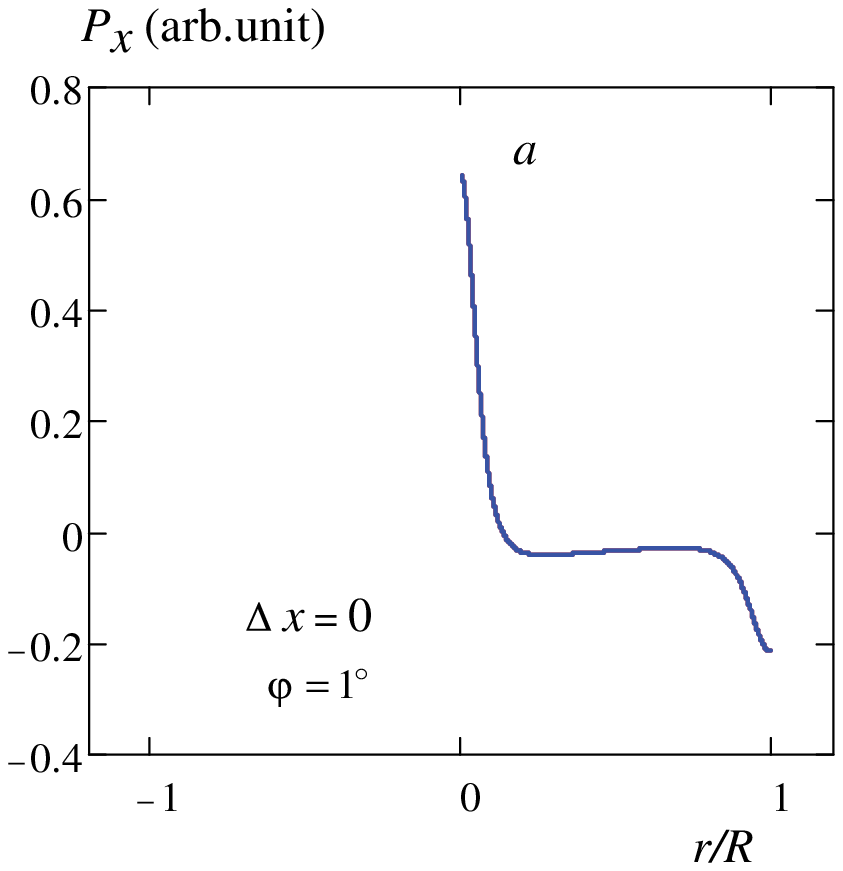}
\includegraphics[width=1.75in,height=1.75in]{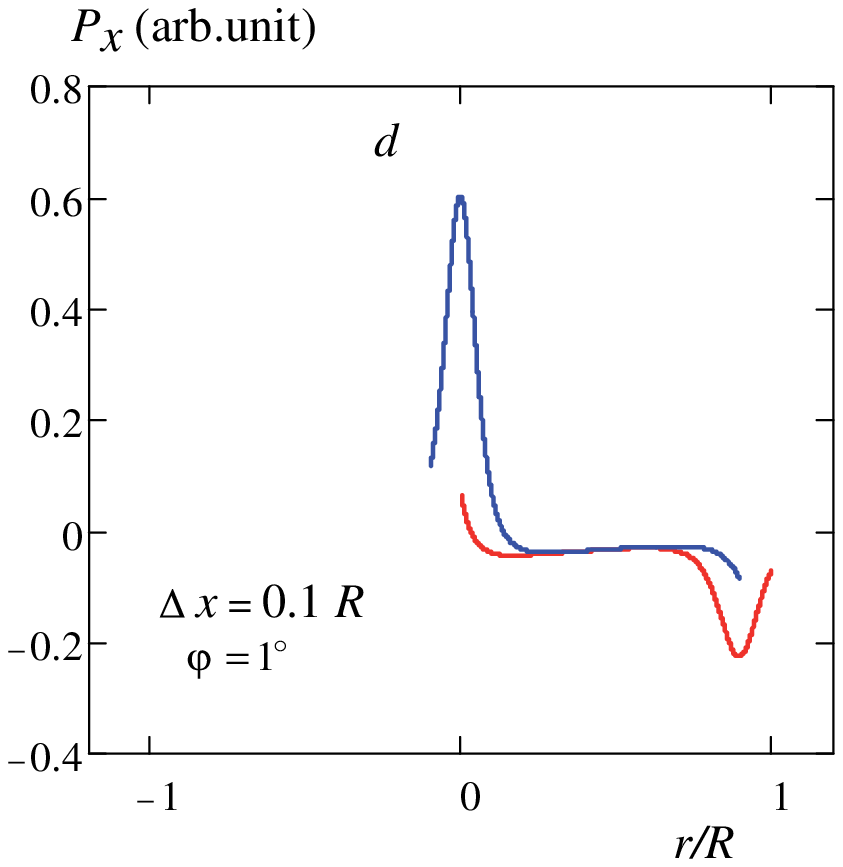}
\includegraphics[width=1.75in,height=1.75in]{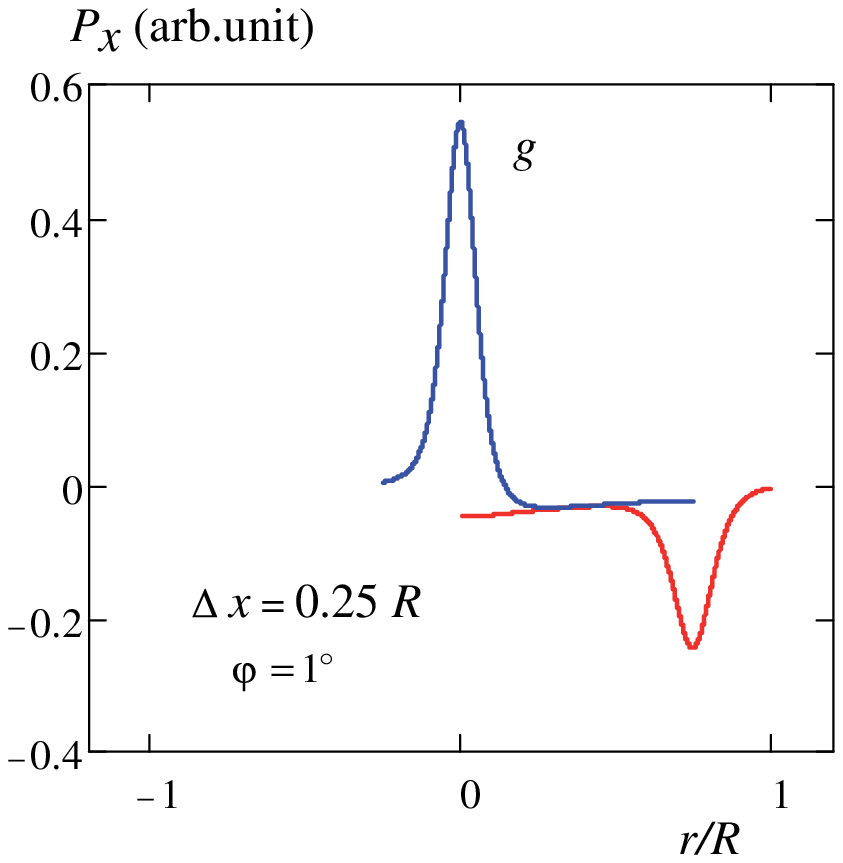}\linebreak }
\centerline{
\includegraphics[width=1.75in,height=1.75in]{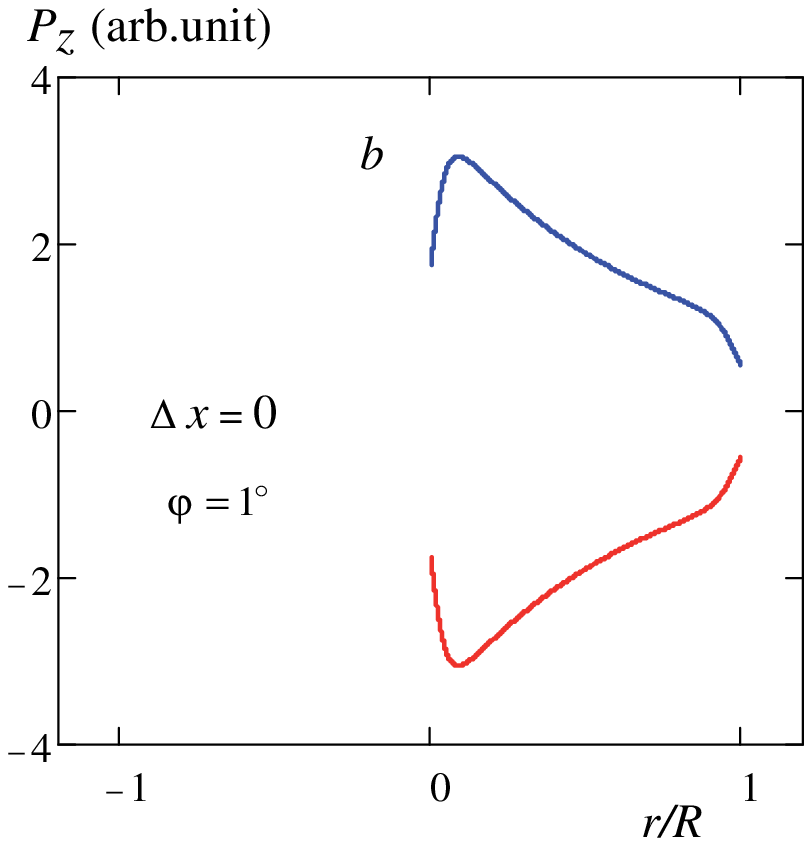}
\includegraphics[width=1.75in,height=1.75in]{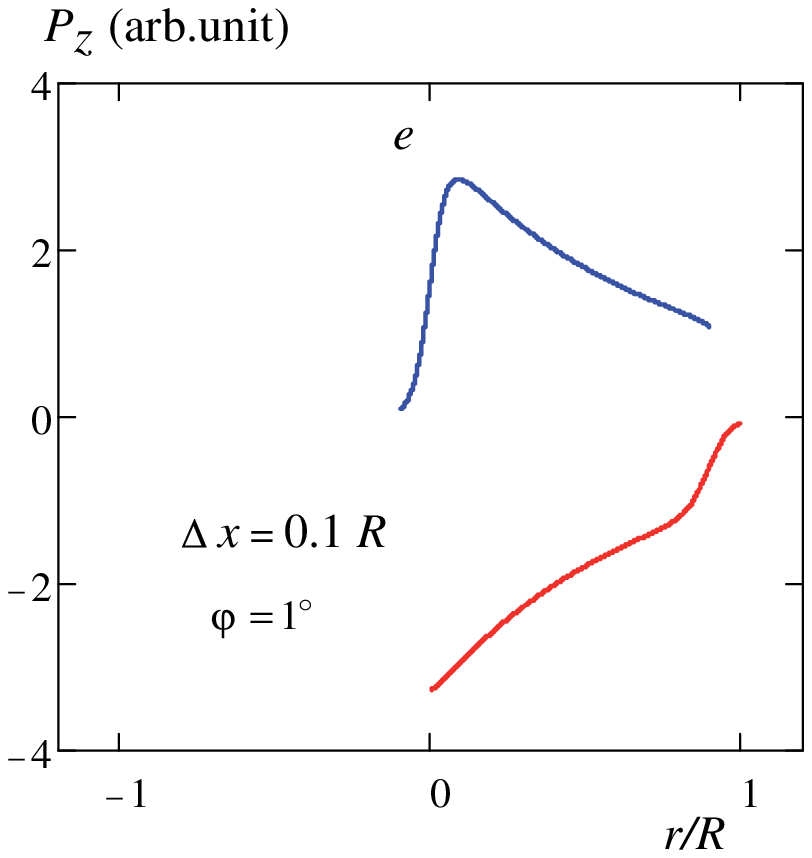}
\includegraphics[width=1.75in,height=1.75in]{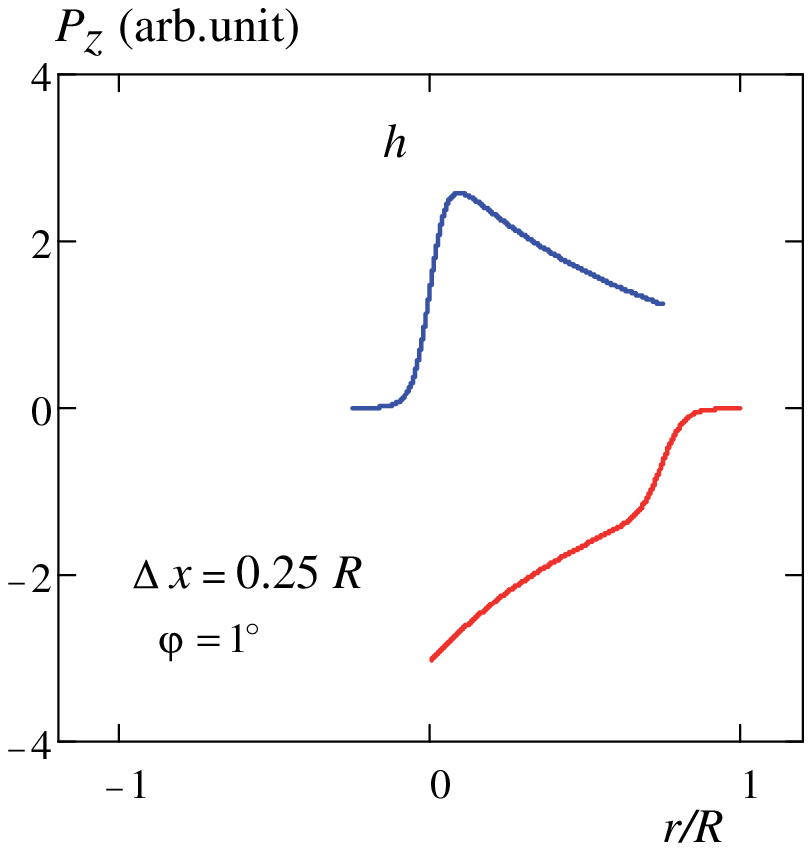}\linebreak }
\centerline{
\includegraphics[width=1.75in,height=0.8in]{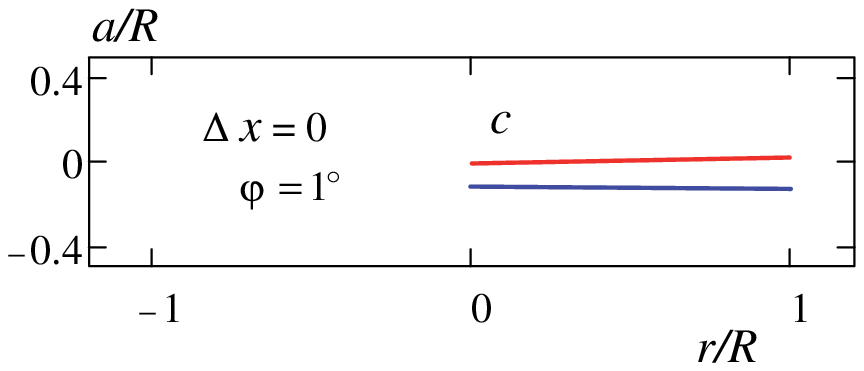}
\includegraphics[width=1.75in,height=0.8in]{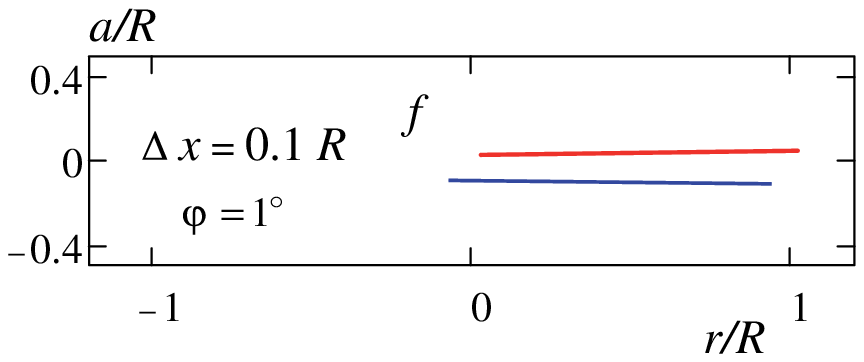}
\includegraphics[width=1.75in,height=0.8in]{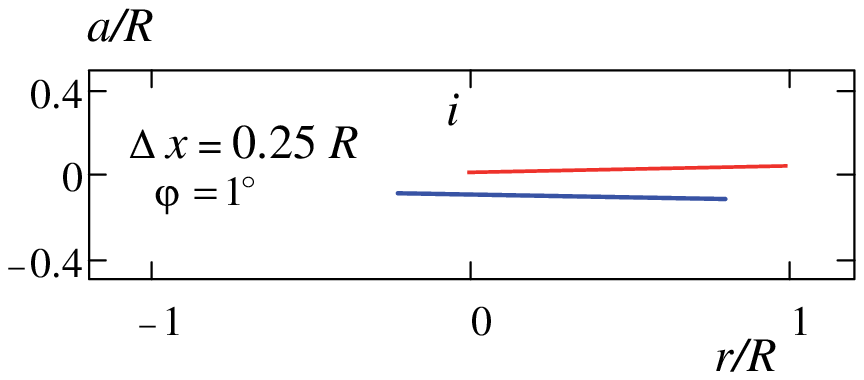}\linebreak }
\label{fig4}
 \caption{Specific local forces of expulsion ($a, d, g)$ and pressure ($b, e, h$) along the $x$ 
axis for the plates ($\varphi =1^\circ $) at different relative shifts $\Delta 
x/R$. In the figure, red lines show the dependences of the local forces 
corresponding to the left plates ($c, f, i$) and the blue lines show the dependences 
corresponding to the right plates.}
\end{figure*}

In Fig.\hyperlink{fig2} 2$c$ it is seen that at the shift of the left plate by the quantity 
$\Delta x = 4\times 10^{-7}\;\mbox{m}$ against the $x$ axis, forces of 
expulsion become uncompensated on the right plate. In this case, the local 
expulsive forces of the right plate in the $x$ direction decrease starting 
from the point $r=0$; however, at the point $(R-\Delta x)/R$ an extremum 
appears. From Fig.\hyperlink{fig2} 2$d$ it also follows that for $r=0$ at the end of the right 
plate at the shift $\Delta x=4\times 10^{-7}\;\mbox{m}$ the Casimir pressure 
grows practically to the values characteristic of points close to $r=R 
\mathord{\left/ {\vphantom {R 2}} \right. \kern-\nulldelimiterspace} 2$.

Obviously, noncompensated Casimir force in the $x$ direction will appear on 
the left plate, which is similar to that found on the right plate (see Fig. \hyperlink{fig3}
3), which is however oppositely directed. It means that at the shift of the 
left plate by $\Delta x$ against the $x$ axis, forces appear in the 
configuration, which tend to bring the plate back to the state before the 
shift to the value $\Delta x=0$. In addition, the same forces will create a 
clockwise torque around the center of mass of the configuration in the plane 
($x,z$) parallel to the $y$ axis. In contrast to the expulsive forces, the 
Casimir pressure along the $z$ axis will have a symmetric form relative to 
the configuration centre at any shift of the plates (Fig.\hyperlink{fig3} 3$b$, $e$, $h$, $k$). 
\begin{figure*}[htbp]
\hypertarget{fig5}
\centerline{
\includegraphics[width=1.6in,height=1.6in]{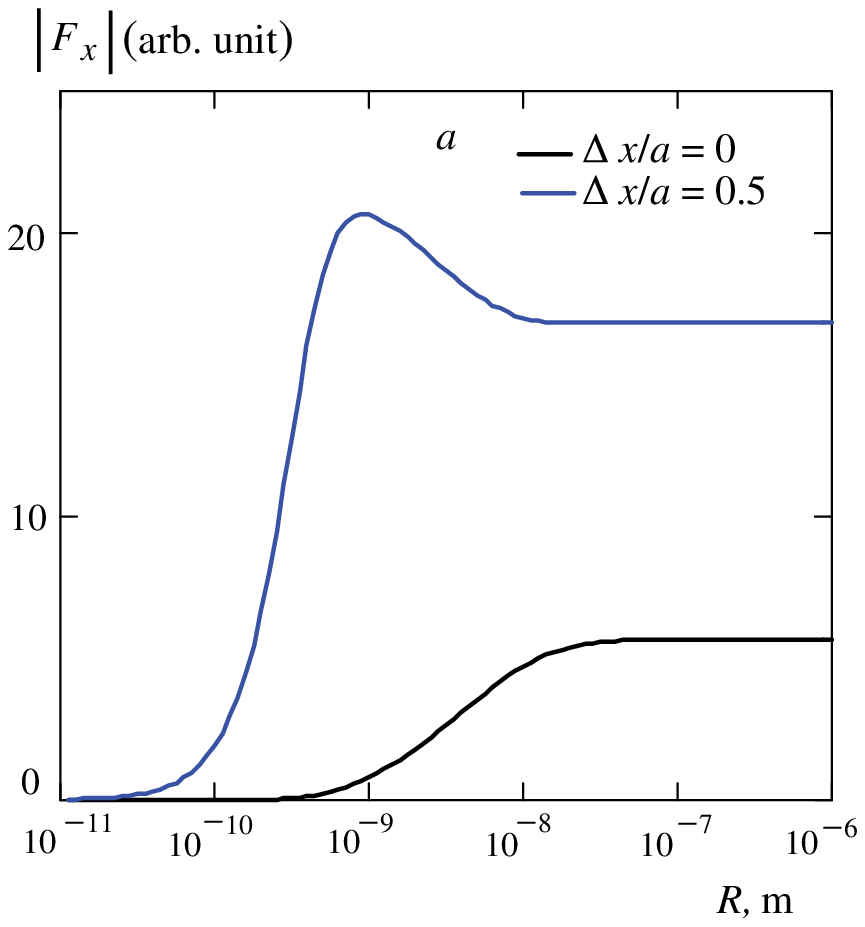}
\includegraphics[width=1.6in,height=1.6in]{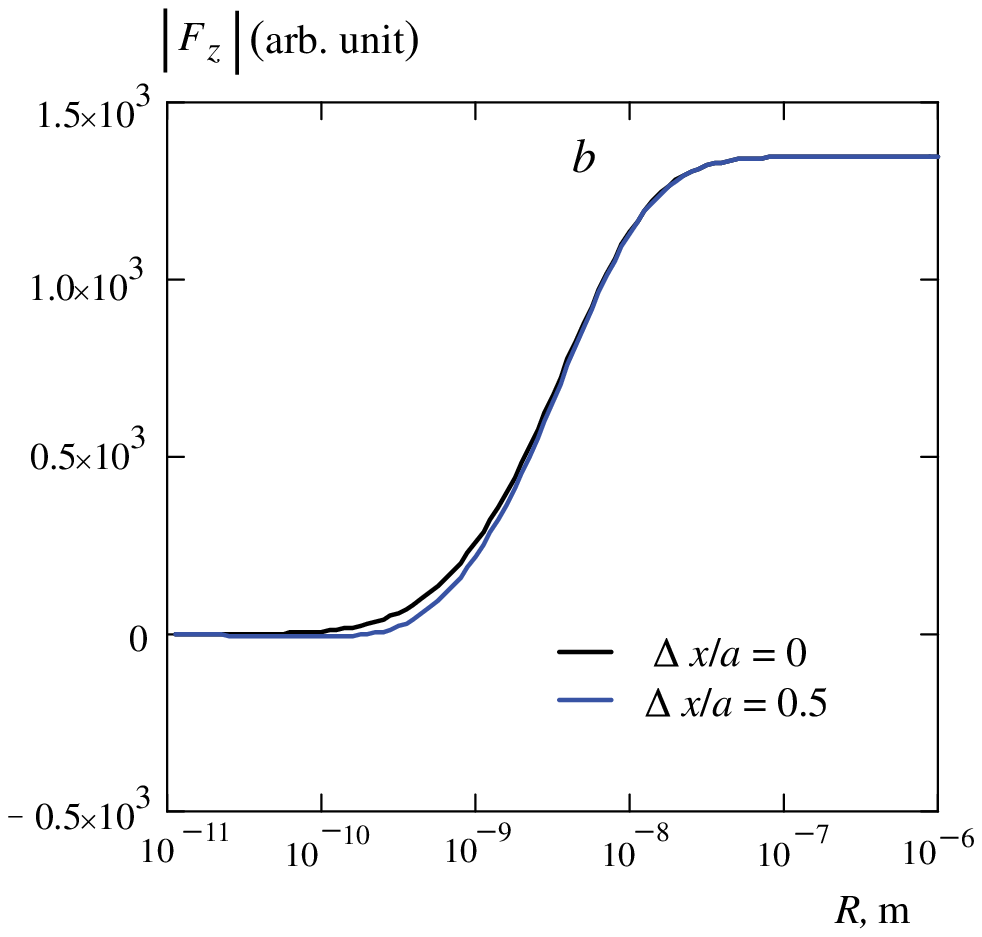}\linebreak }
\centerline{
\includegraphics[width=1.6in,height=1.6in]{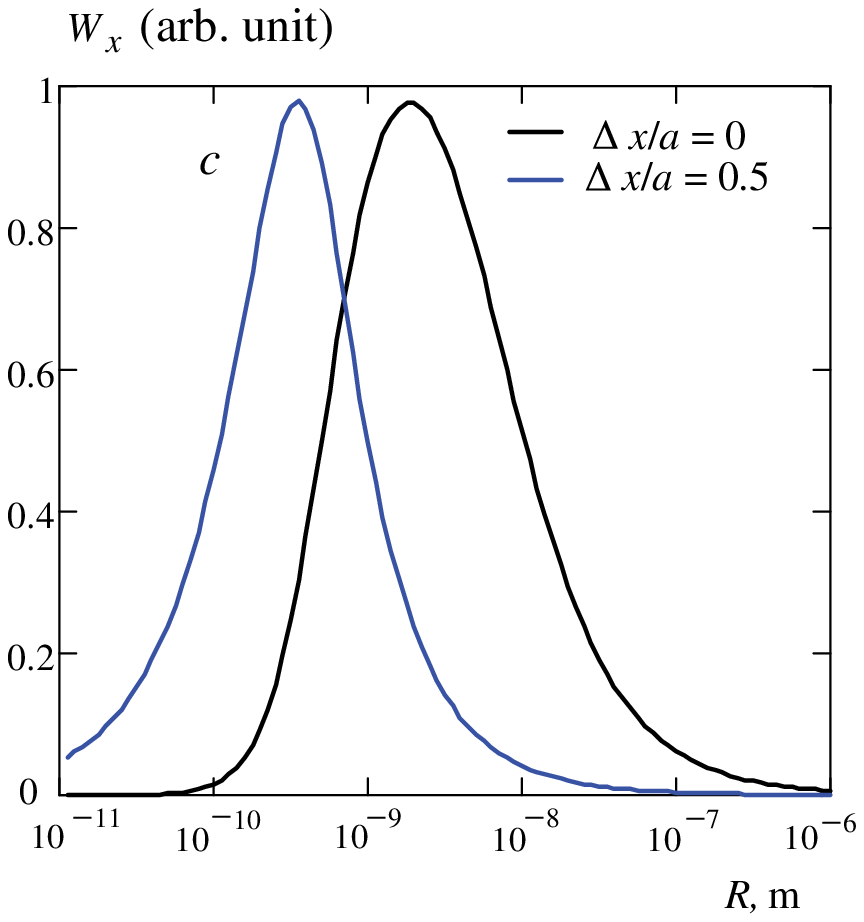}
\includegraphics[width=1.6in,height=1.6in]{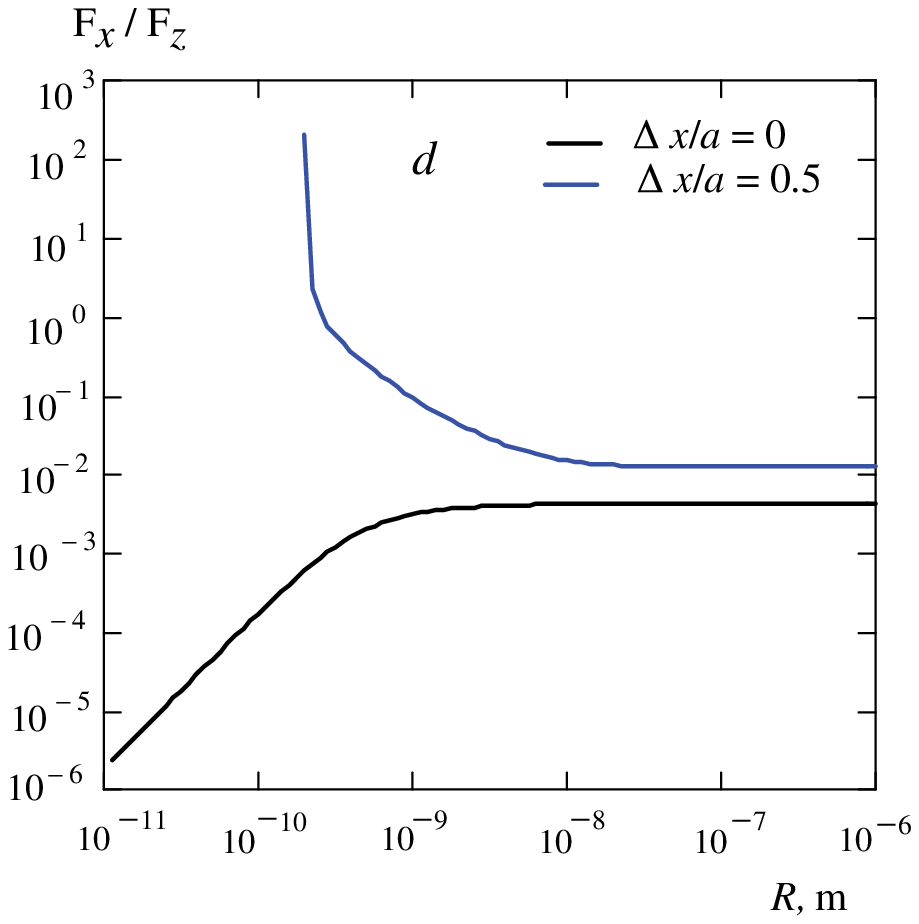}\linebreak }
\label{fig5}
\caption{The total Casimir force of expulsion ($a$) and compression ($b$) of the 
cavity wing at $\varphi =1\mathring{ }$ depending on the wing length $R$ and at 
the $\Delta x/R$ shift of the left plate relative to the right one. The 
dependence of the effectiveness of the expulsion $W_x$ on the wing length 
($c$) at the shift. The relation of the total forces of expulsion to the forces 
of compression $F_x /F_z $ depending on the wing length $R$ ($d$).}
\end{figure*}
\begin{figure*}[htbp]
\hypertarget{fig6}
\centerline{
\includegraphics[width=1.6in,height=1.6in]{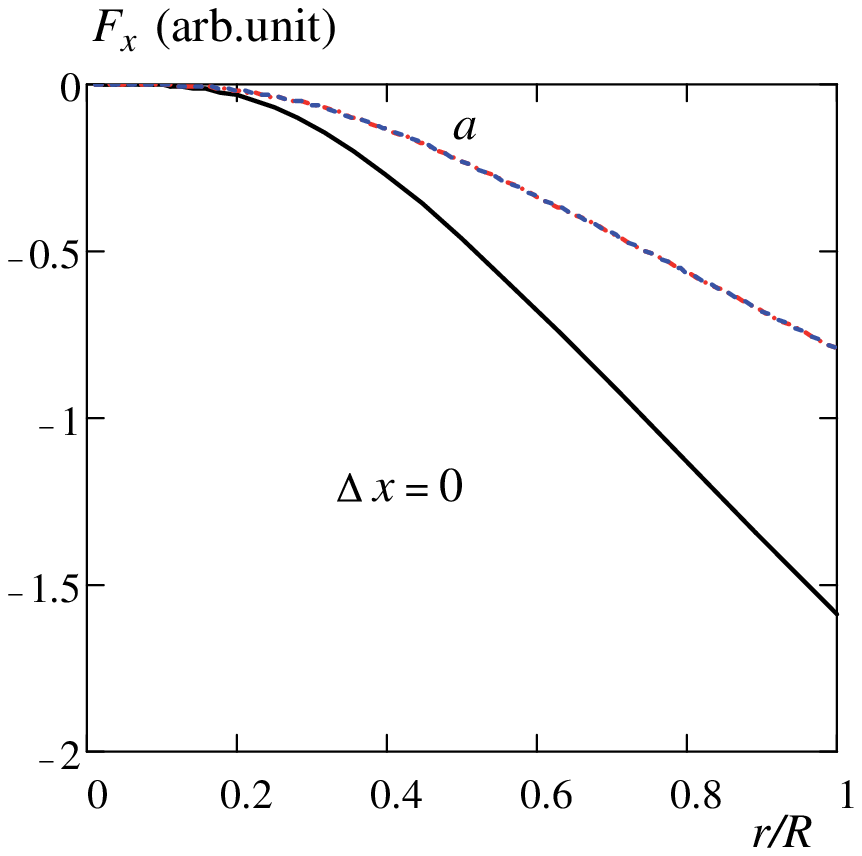}
\includegraphics[width=1.6in,height=1.6in]{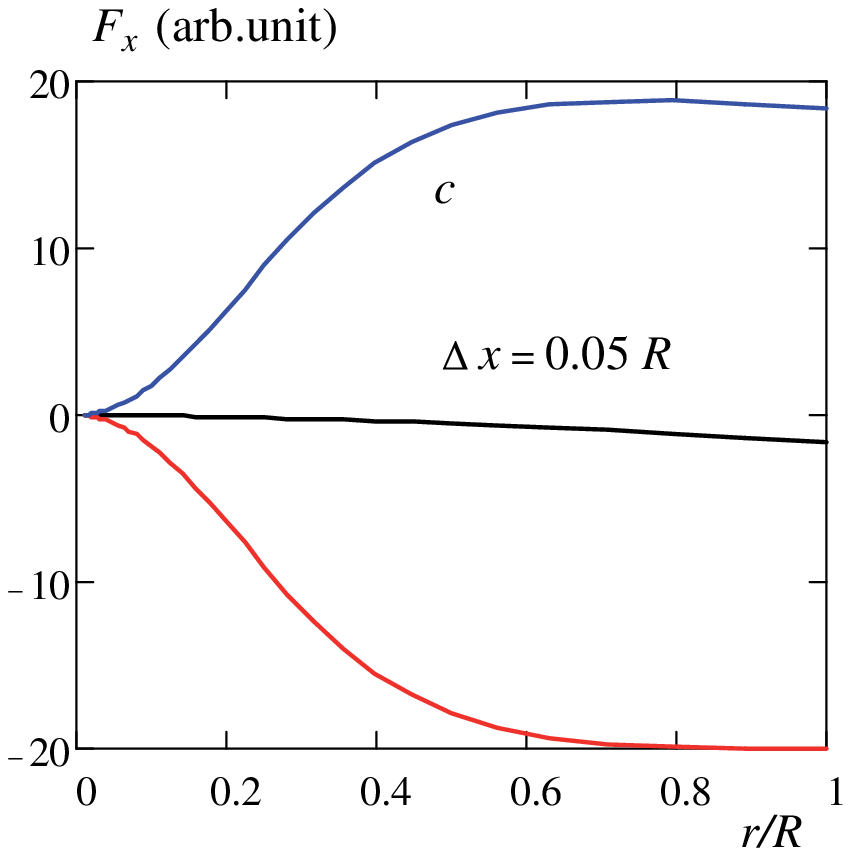}
\includegraphics[width=1.6in,height=1.6in]{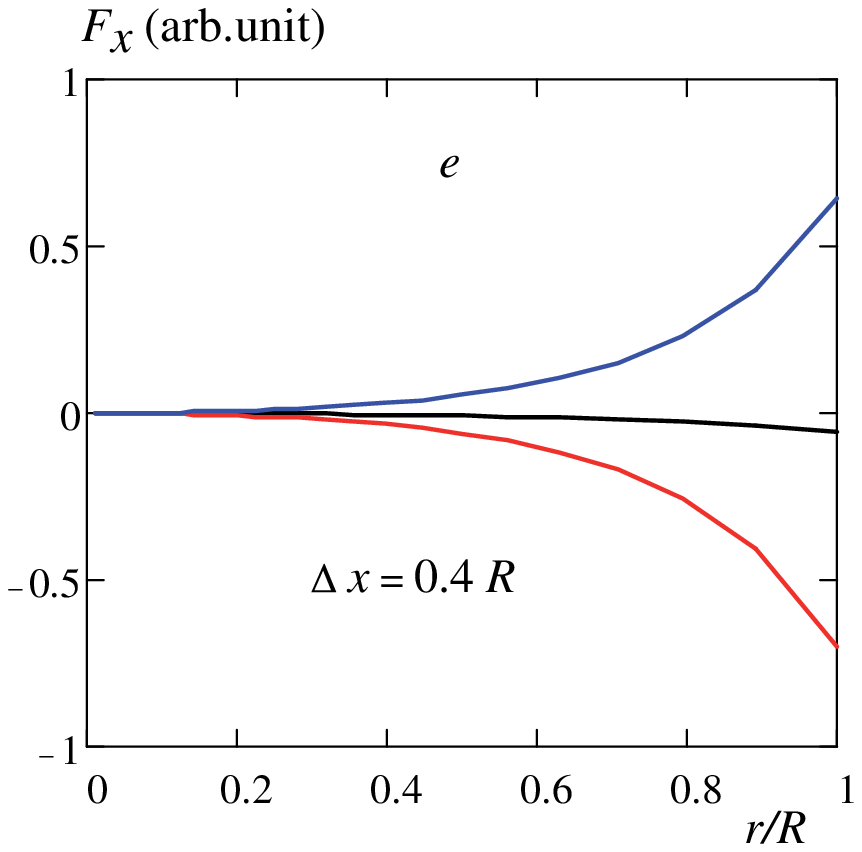}
\includegraphics[width=1.6in,height=1.6in]{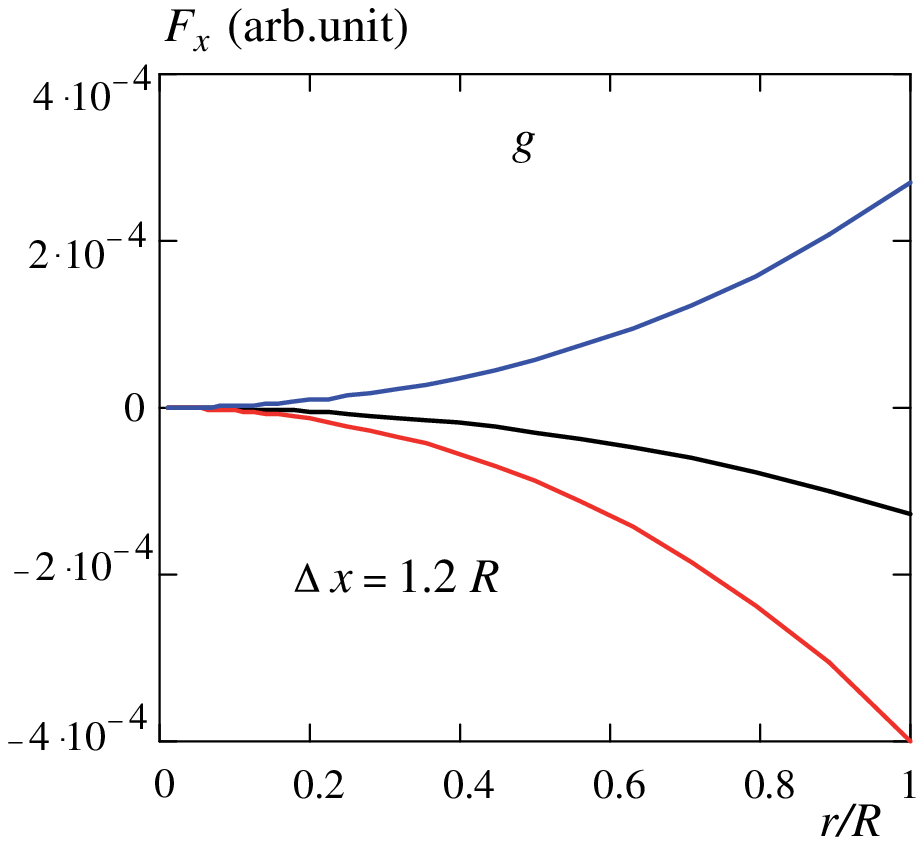}\linebreak }
\centerline{
\includegraphics[width=1.6in,height=0.7in]{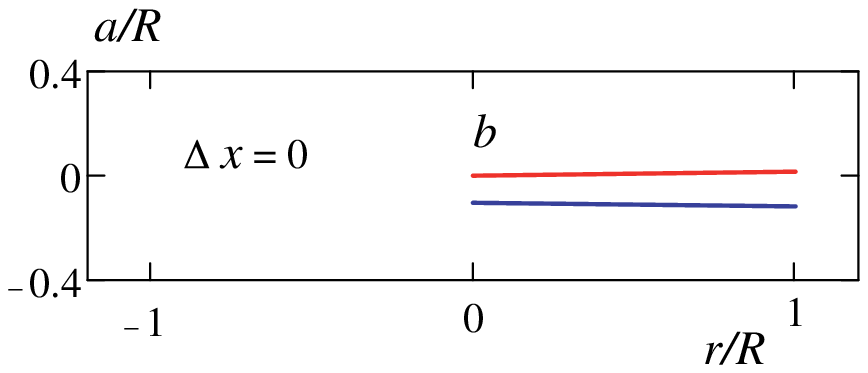}
\includegraphics[width=1.6in,height=0.7in]{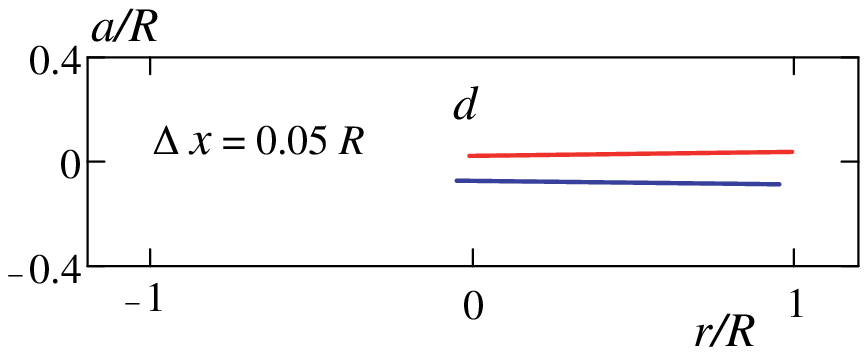}
\includegraphics[width=1.6in,height=0.7in]{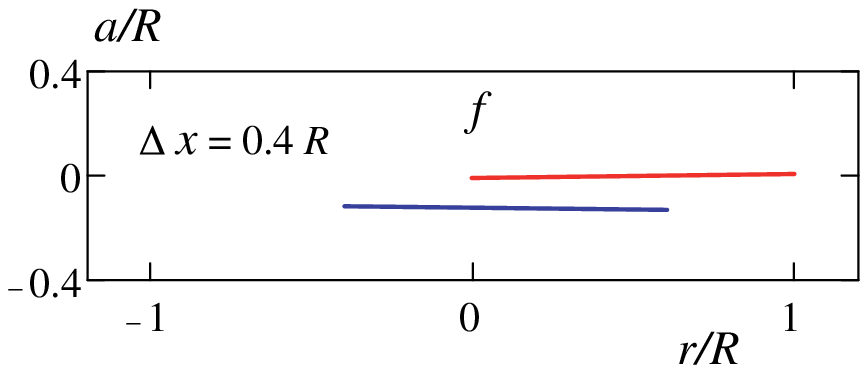}
\includegraphics[width=1.6in,height=0.7in]{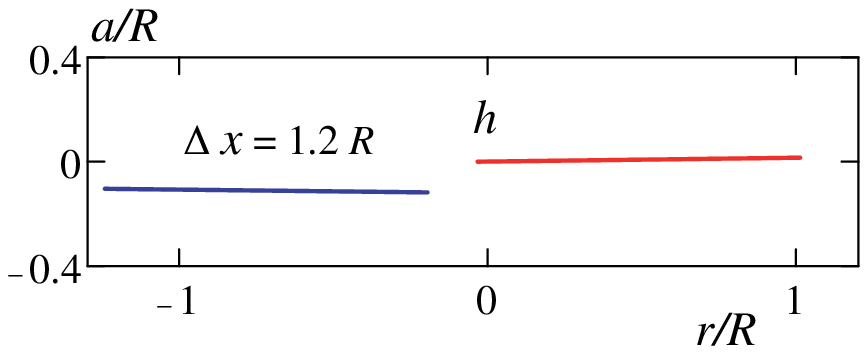}\linebreak }
\label{fig6}
\caption{The Casimir force of expulsion of the entire configuration of the 
cavity wings (\textit{a, c, e, g}) fixed after the shift for $\varphi =1\mathring{ }$ depending on 
the length $r$of the wings and the shift $\Delta x/R$ of the left plate (\textit{b, d, f. h}) 
(red line) relative to the right plate (blue line). The black line shows the 
integral force of expulsion of the configuration fixed after the shift.}
\end{figure*}
\begin{figure*}[htbp]
\hypertarget{fig7}
\centerline{
\includegraphics[width=1.8in,height=1.8in]{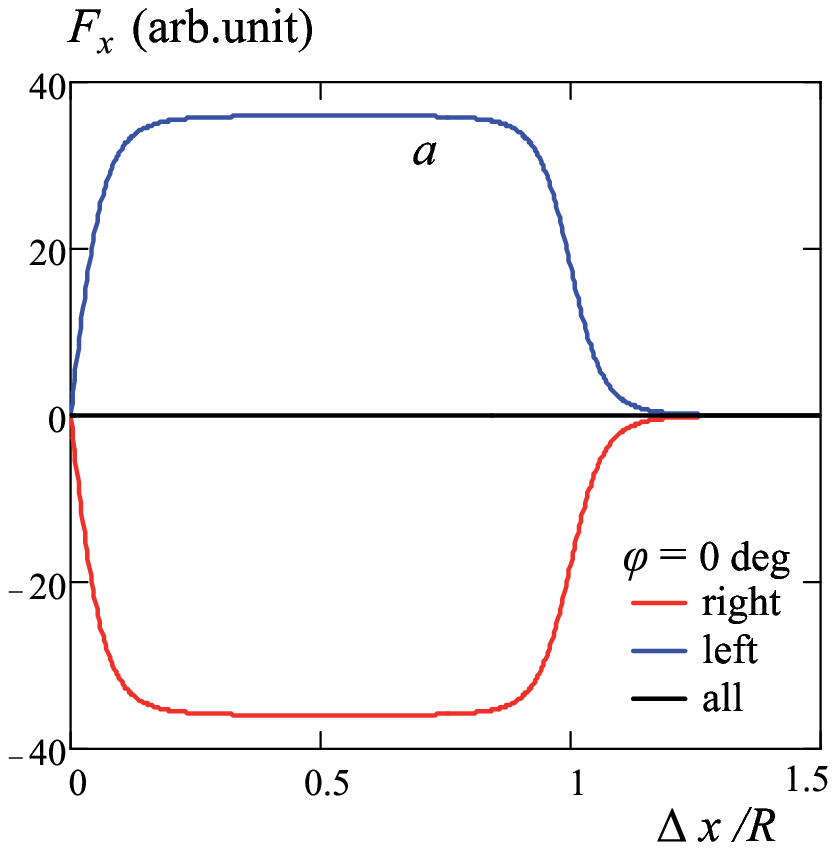}
\includegraphics[width=1.8in,height=1.8in]{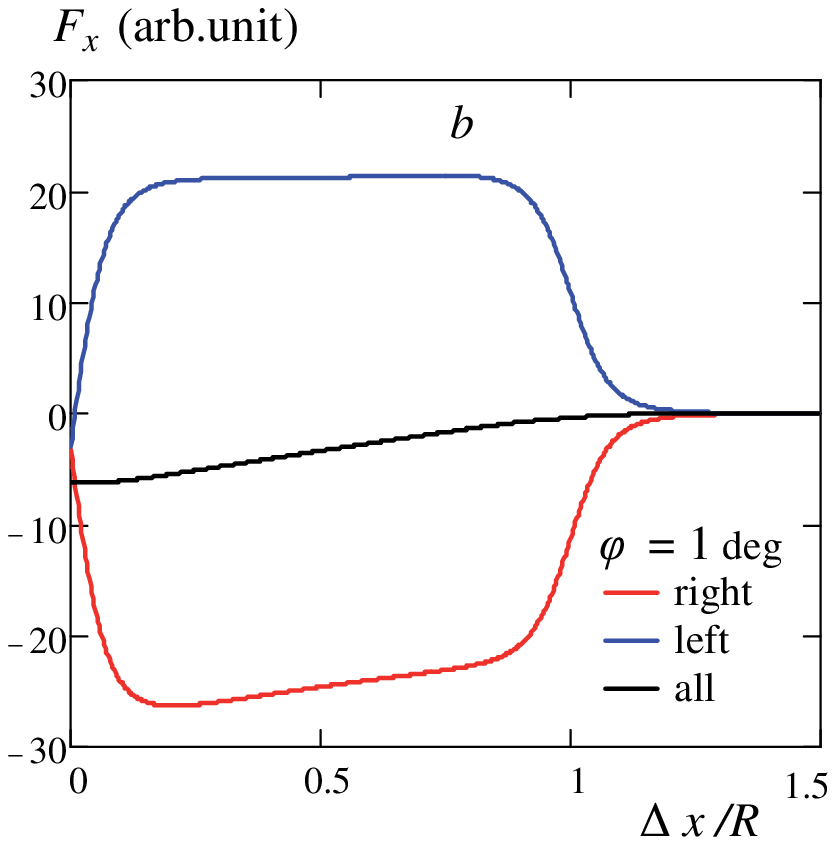}
\includegraphics[width=1.8in,height=1.8in]{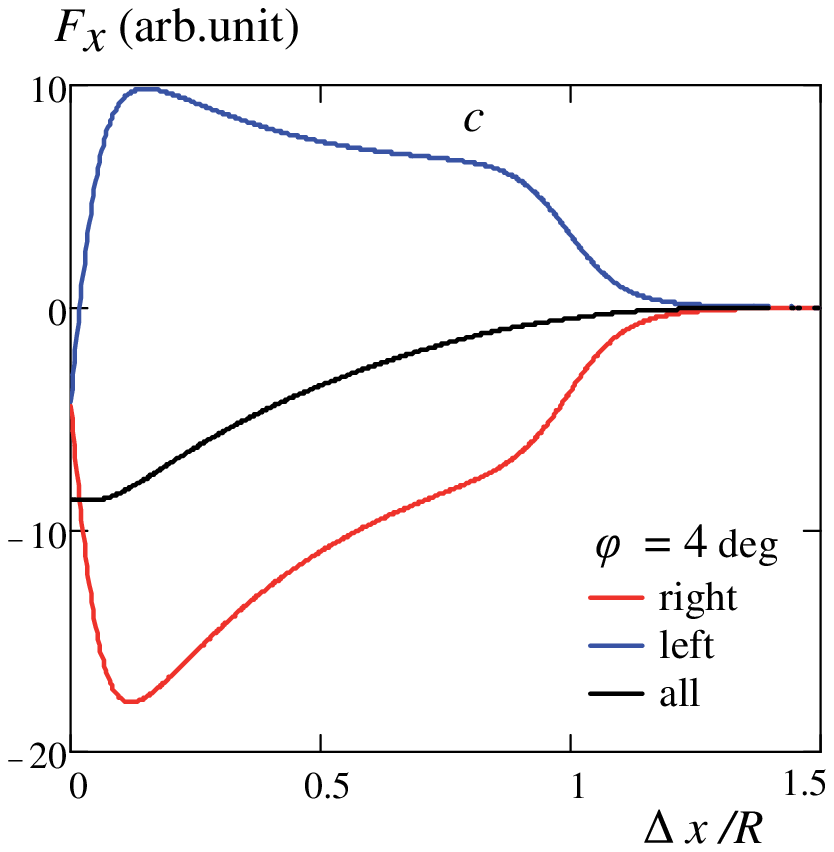}}
\label{fig7}
\caption{The dependences of the forces of expulsion of the right $F_x^{right}$
wing and the left $F_x^{left}$ wing and the sum of these forces $F_x $ on 
the value of the shift $\Delta x/R$ for three angles of the opening of the 
generetrices of the wings.}
\end{figure*}
Let us suppose that after shift the plates will be fixed; in this case in 
the system the clockwise torque will appear in the plane ($x,z$) due to the 
Casimir forces of expulsion. However, if the system is not fixed along the 
$x$ axis after shift and let us suppose that the distance $a$ remains 
constant, the appearance of rather complex oscillatory motions can be 
expected in the system. The oscillations of such system can have the 
following character. After the shift by $\Delta x$, both plates will tend to 
return to the previous state when $\Delta x=0$ and simultaneously will spin 
together in the clockwise direction in the plane ($x,z$). Obviously, the 
plates will return to the opposite state with the shift $-\Delta x$ along 
the $x$ axis. When imperfect systems are considered, it is necessary to take 
into account the processes of dissipation in them. In this case the system 
will return to the state which is not absolutely opposite with the shift by 
$-\Delta x$ relative to the $x$ axis. Thus, the system of the plates will 
tend in the direction opposite to the shift motion and simultaneously it 
will spin in the opposite (anticlockwise) direction. That is in the unfixed 
system of plates, after the shift of the plates relative to one another 
complex cyclic processes can be observed.

When the cavity wings are open up to the permissible limit angle $\varphi 
\leqslant \mbox{arccot}\left( {{\Delta x} \mathord{\left/ {\vphantom 
{{\Delta x} a}} \right. \kern-\nulldelimiterspace} a} \right)$, with growing 
$r/R\to 1$ the specific forces are decreasing both along the $x$ axis and 
along the $z$ axis (Fig.\hyperlink{fig4} 4$a$, $b$). However, the character of the spinning action 
of the forces on the plates weakens with the growth of the angle $\varphi $ and 
the direction of the shifting forces remains similar to that which is 
observed in the case of parallel plates. Such character of the weakening of 
the forces at the shift of planes and the simultaneous increase of the angle 
of their opening $\varphi $ significantly differs from that observed in the 
situation when there is no shift (see Fig.\hyperlink{fig3} 3).

By integrating $P_x (r)$ and $P_z (r)$ with respect to $x$ we find the total 
Casimir force $F_x $ and $F_z $ acting on the right wing of the cavity at 
the shift of the left wing. Let us investigate the dependence of the Casimir 
forces on the wing length $R$. Also, let us find the most effective wing 
length at which there is the maximum of the function and the relation $F_x 
/F_z $. The corresponding results are displayed in Fig.\hyperlink{fig5} 5.

As seen in Fig. \hyperlink{fig5}5$a$, the total expulsive force of the shifted configuration is 
always directed against the $x$ axis similar to the case of the 
configuration without shift. The force manifests itself as time-constant 
expulsion of opened trapezoid cavity in the direction of its least opening 
(i.e. in the direction of the smaller section). The forces grow with the 
growth of the cavity wing length to a certain best length $R_{eff}$ (Fig.\hyperlink{fig5} 
5$c$), after which the growth of the forces continue but the relation of the 
forces to the wing length decreases. It means that the expulsion force is 
most effectively manifested in the region limited by the size $R_{eff}$. 
For the configurations without shifts, at the angle $\varphi =1\mathring{ }$ 
this optimum has the theoretical value $R_{eff} =1.85\times 10^{-9}$ \,m (1.85 
nm) and, correspondingly, $F_x \sim 5.7\,\mbox{N}$ for one cavity wing with 
$R = R_{eff}$ and the length $l=1$ \,m. However, at a small shift of the left 
plate ($\Delta x/a=0.5)$ the entire force of the figure expulsion grows to 
the values $F_x \sim 21\,\mbox{N}$. Such shift in the configuration leads to 
a considerable decrease in the theoretical value $R_{eff}$ down to $R_{eff} 
\sim a$ (0.4 nm). Here the values of forces are given in real quantities.
\begin{figure*}
\hypertarget{fig8}
\centerline{
\includegraphics[width=1.8in,height=1.8in]{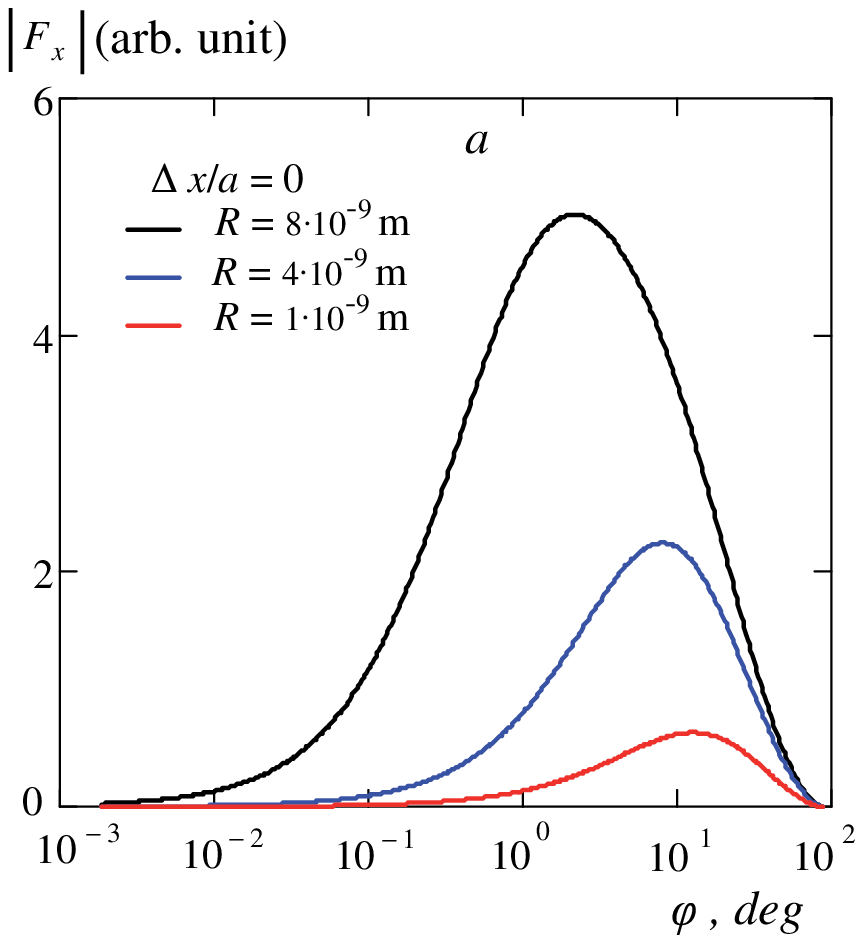}
\includegraphics[width=1.8in,height=1.8in]{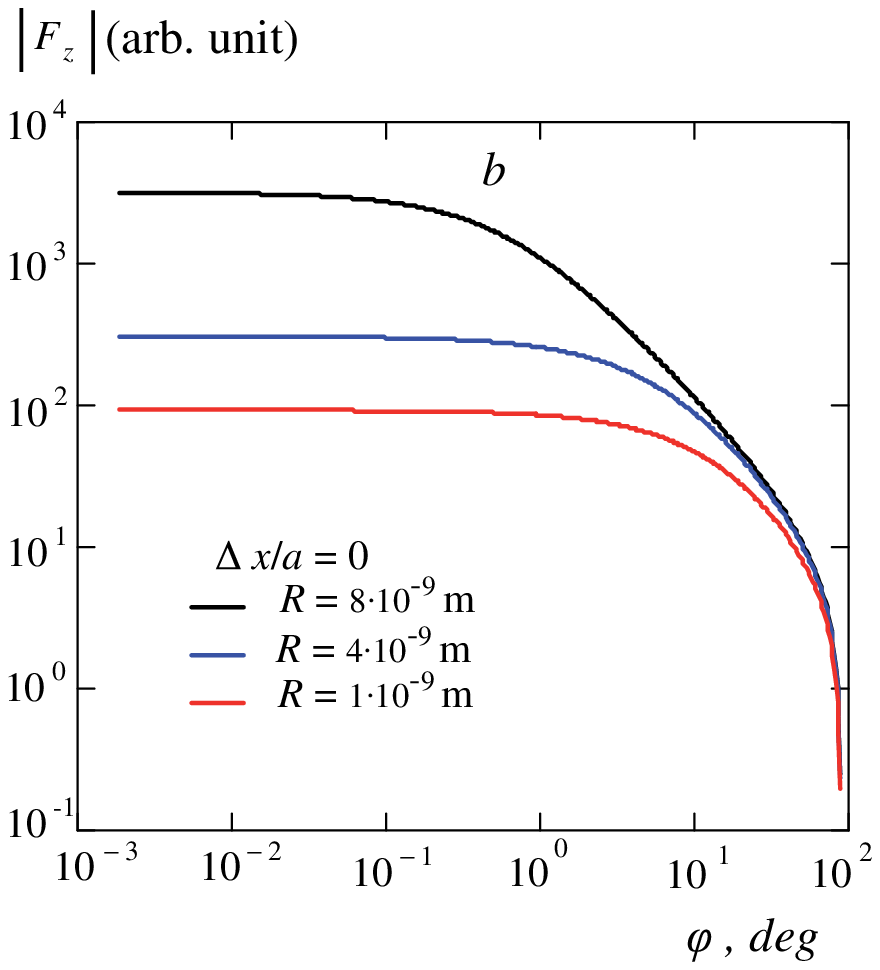}
\includegraphics[width=1.8in,height=1.8in]{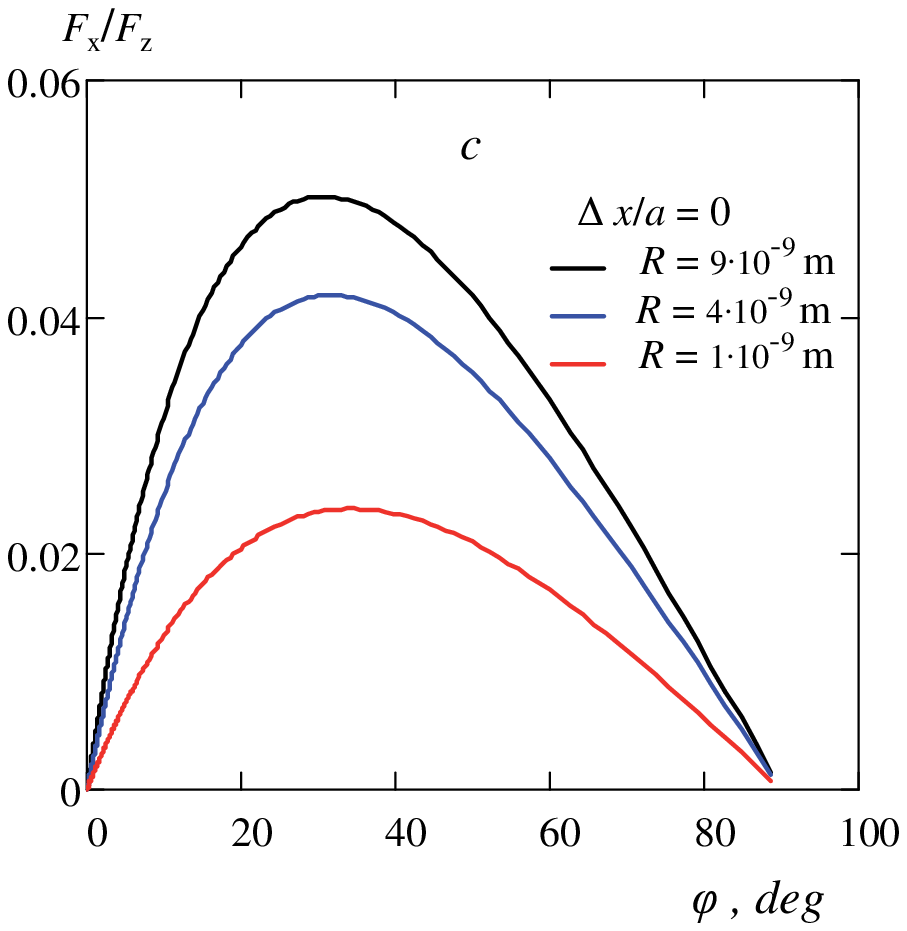}\linebreak }
\centerline{
\includegraphics[width=1.8in,height=1.8in]{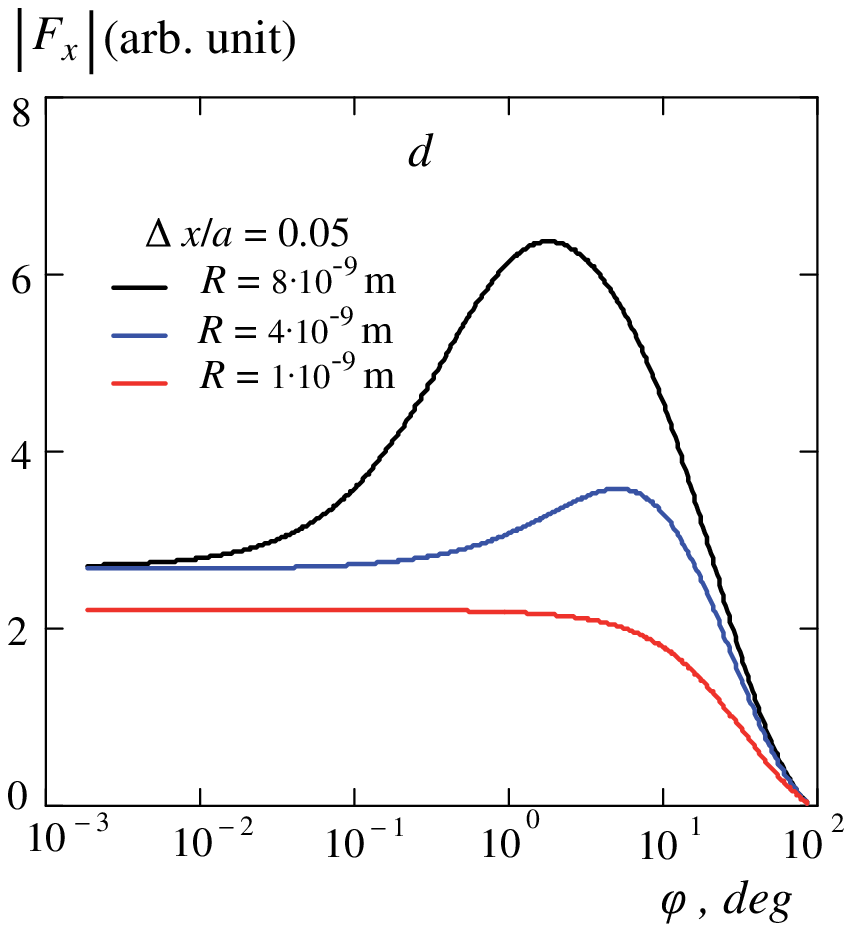}
\includegraphics[width=1.8in,height=1.8in]{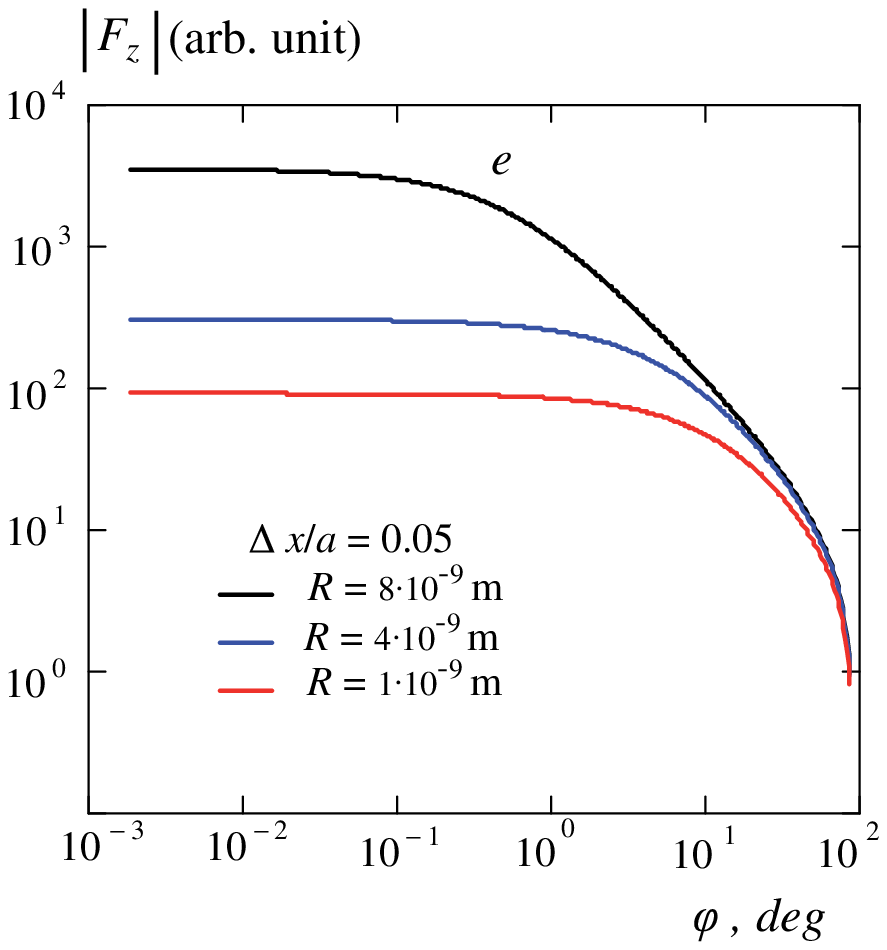}
\includegraphics[width=1.8in,height=1.8in]{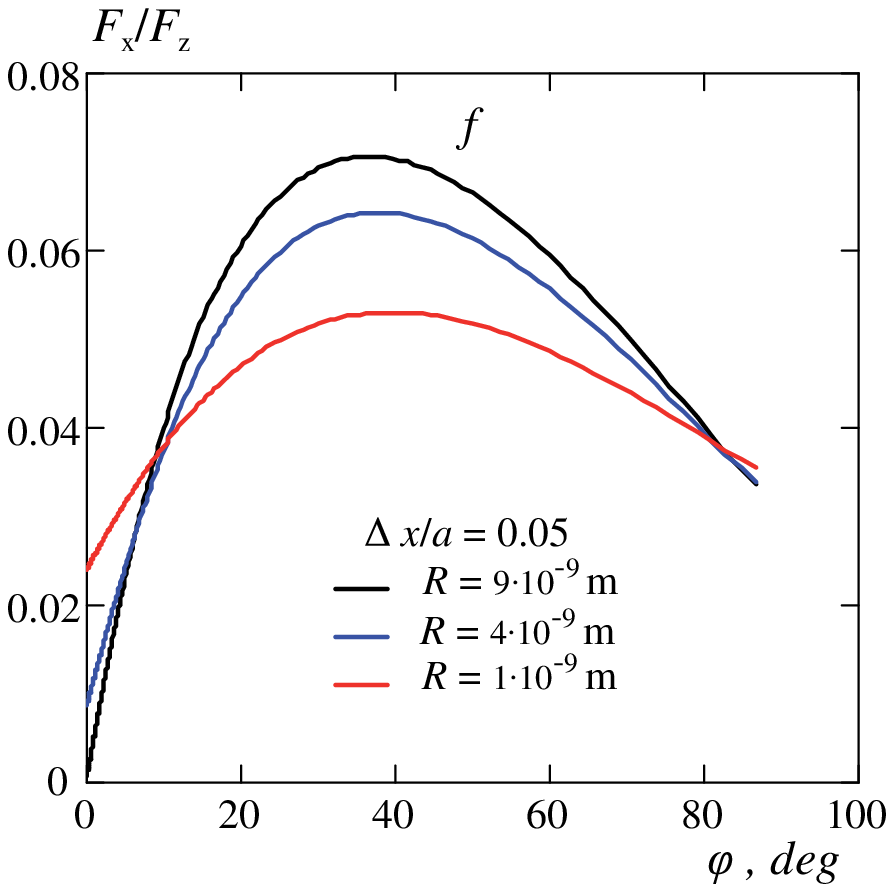}\linebreak }
\centerline{
\includegraphics[width=1.8in,height=1.8in]{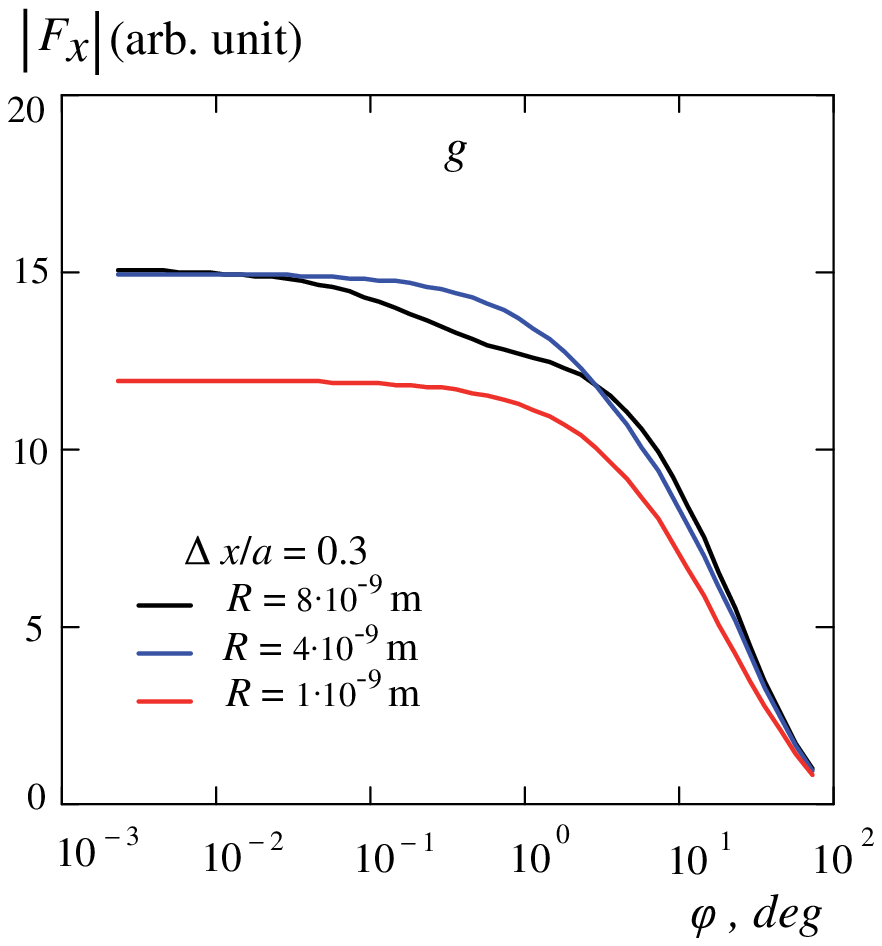}
\includegraphics[width=1.8in,height=1.9in]{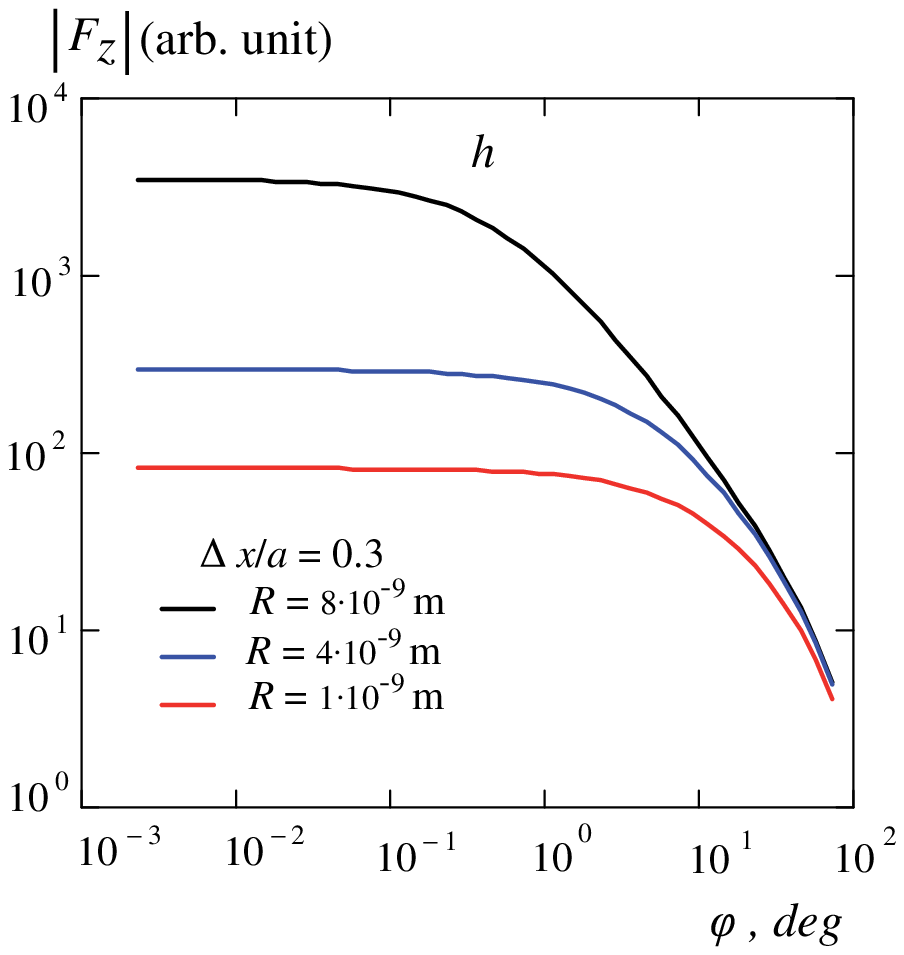}
\includegraphics[width=1.8in,height=1.8in]{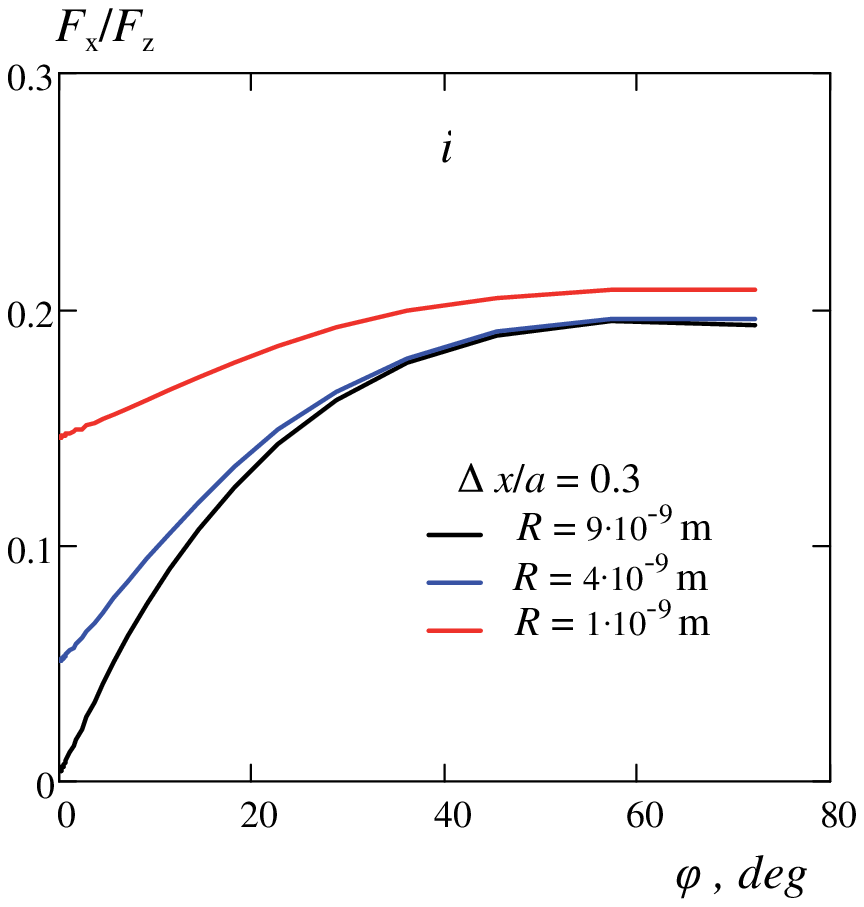}\linebreak }
\label{fig8}
 \caption{Absolute values of the Casimir forces of expulsion ($a$) and 
compression ($b$) for different lengths $R$ depending on the angle $\varphi $ for 
the configurations without shift $\Delta x/a=0$. ($c$) -- the relation of the 
total forces of expulsion to the compression depending on the angle $\varphi $. 
The corresponding forces and relations for the configuration with the shift 
$\Delta x/a=0.05$ ($d, e, f)$ and $\Delta x/a=0.3_{ }(g, h, i)$.$_{ }$}
\end{figure*}

Note that the shift does not influence the character of the growth of the 
Casimir pressure at the growth of the cavity wing length $R$. The relation 
of total expulsive forces to the compressive forces $F_x /F_z $ acquires 
radically different character depending on the wing length $R$ at the shift 
in the configuration (Fig. \hyperlink{fig5}5$d$). Instead of gradual growth of the value 
$\left| {F_x /F_z } \right|$ up to $\left| {F_x /F_z } \right|\to 4.2\times 
10^{-3}$, at the shift of the plates by $\Delta x/a=0.5$ the theoretical 
relation $\vert F_{x} / F_{z}\vert $ decreases from the value $\left| {F_x 
/F_z } \right|\sim 10^2$ to $\left| {F_x /F_z } \right|\to 1.3\times 
10^{-2}$. For the value $R_{eff} \sim a$ the relation is $\left| {F_x /F_z } 
\right|\approx 0.46$.

The Casimir forces of expulsion for the entire configuration (if after the 
shift of the wings the configuration is fixed in the new position) with the 
angle of opening $\varphi =1\mathring{ }$ at different $\Delta x$ are shown in 
Fig.\hyperlink{fig6} 6 depending on the length of the wings. In the absence of shift 
($\Delta x=0)$ both wings expulsed against the $x$ direction (Fig. \hyperlink{fig6}6$a$, $b$). If 
the wings are rigidly connected the configuration is expulsed against the 
$x$ axis in two times stronger. Even a very small shift of the left plate 
relative to the right plate (for example, by $\Delta x/R=0.05$) leads to an 
abrupt change in the direction and value of the expulsive forces. The left 
wing is expulsed in the $x$ direction and the right wing against the $x$ 
axis with the force larger by a factor of 10 than that in the absence of 
shift (Fig.\hyperlink{fig6} 6$c$, $d)$. In this case the integral force of the expulsion of the 
fixed configuration is directed against the $x$ axis, and its value makes $\sim 
99\% $ of that of the force before the shift. At the shift of the order of 
$\Delta x/R=0.4$ the system of two connected plates is also expulsed against 
the $x$ axis but much weaker (Fig.\hyperlink{fig6}6$e$, $f$) than in the case of smaller shifts.

At $\Delta x/R=1.2$ and larger shifts (by $\Delta x/R\gg 1$) the integral 
forces of expulsion retain the direction but rapidly become weaker to 0. At 
the increase of the angle $\varphi $ the direction of the expulsive forces 
remains the same but the integral forces become even weaker. This also 
follows from the shift value $\Delta x/R$ dependence of the forces of 
expulsion of the right $F_x^{right} $ and left $F_x^{left} $ wings and the 
sum of these forces $F_x =F_x^{right} +F_x^{left} $ in the configuration 
fixed after the shit (Fig.\hyperlink{fig7} 7). 

Figure 8$a,b$ shows the dependences of the integral forces of expulsion and 
compression of the configuration right wing on the angle $\varphi $. It can be 
seen that in the absence of shift for any length $R$ of the cavity wing 
$\Delta x/a=0$ there is the maximum of compulsion depending on the angle 
$\varphi $. The larger is the length $R$, the smaller is the angle. Fig. \hyperlink{fig8}8$c$ 
presents the relation $F_x /F_z $ depending on the angle $\varphi $. 

Even though the shift of the configuration is $\Delta x/a=0.05$, a 
significant change in the character of the angle $\varphi $ dependences of the 
forces of expulsion takes place (Fig. \hyperlink{fig8}8$d$). However, the character of the 
Casimir along the $x$ axis does not practically change at the configuration 
shift (Fig.\hyperlink{fig8} 8$e$, $h$). When the shift by $\Delta x/a=0.3$ takes place, a 
considerable growth of expulsive forces is observed for very small angles 
$\varphi $ of the opening of cavities. Most clearly it can be observed for the 
lengths $R_{eff} \leqslant a$ (Fig.\hyperlink{fig8} 8$g$).
\begin{center}
\textbf{Conclusions}
\end{center}

In the present paper, it is shown that the shifts in configurations of 
trapezoid metal (perfectly conducting) nanosized cavities lead to 
significant changes in the character of dependences of the noncompensated 
Casimir forces (forces of expulsion) on the length of wings and the angles 
of their opening. At such shifts the character of the change in the Casimir 
pressure does not practically depend on the wings' length and the angles of 
their opening. In the case of parallel metal mirrors, it is found that after 
the shift in both plates the appeared forces of expulsion are directed so 
that they are tending to return the system in the state before the shift. In 
the case of the unfixed arrangement of plates relative one another at the 
fixed distance, after shifts complex oscillatory states with torques can 
appear in the system of the plates. In the case of the arrangement of plates 
the direction of expulsion can change in configurations depending on the 
degree of the shift of their elements. At any shift of elements a 
time-constant torque can appear in a configuration. 
\begin{acknowledgments}
The author is grateful to T. Bakitskaya for hers helpful
participation in discussions.
\end{acknowledgments}

\end{document}